\documentclass[11pt]{article}

\usepackage{euscript}
\usepackage{amssymb}
\usepackage{amsfonts}
\usepackage{amsbsy}
\usepackage{amsmath}
\usepackage{epsfig}
\usepackage{amsthm}
\usepackage{amscd}
\usepackage{amstext}
\usepackage{verbatim}

\usepackage[pdftex]{hyperref}

\def\ben{\begin{equation}}
\def\een{\end{equation}}

\def\be{\begin{equation}}
\def\ee{\end{equation}}
\def\beq{\begin{equation}}
\def\eeq{\end{equation}}
\def\ba{\begin{array}}
\def\ea{\end{array}}

\def\dalemb#1#2{{\vbox{\hrule height .#2pt
       \hbox{\vrule width.#2pt height#1pt \kern#1pt
               \vrule width.#2pt}
       \hrule height.#2pt}}}

\newcommand{\bea}{\begin{eqnarray}}
\newcommand{\eea}{\end{eqnarray}}

\numberwithin{equation}{section}

\begin{document}

\begin{center}

{ \LARGE {\bf Warped Conformal Field Theory as Lower Spin Gravity}}

\vspace{1.2cm}

Diego M. Hofman$^\sharp$ and Blaise Rollier$^\flat$ 

\vspace{0.9cm}

{\it $^\sharp$ Institute for Theoretical Physics, University of Amsterdam,\\
Science Park 904, Postbus 94485, 1090 GL Amsterdam, The Netherlands \\}

\vspace{0.5cm}

{\it $^\flat$ Van Swinderen Institute for Particle Physics and Gravity, University of Groningen, \\
Nijenborgh 4, 9747 AG Groningen, The Netherlands \\}

\vspace{1.6cm}

{\tt d.m.hofman@uva.nl, B.R.Rollier@rug.nl} \\

\vspace{1.6cm}

\end{center}

\begin{abstract}
Two dimensional Warped Conformal Field Theories (WCFTs) may represent the simplest examples of field theories without Lorentz invariance that can be described holographically. As such they constitute a natural window into holography in non $AdS$ space-times, including the near horizon geometry of generic extremal black holes. It is shown in this paper that WCFTs posses a type of boost symmetry. Using this insight, we discuss how to couple these theories to background geometry. This geometry is not Riemannian. We call it Warped Geometry and it turns out to be a variant of a Newton-Cartan structure with additional scaling symmetries. With this formalism the equivalent of Weyl invariance in these theories is presented and we write two explicit examples of WCFTs. These are free fermionic theories. Lastly we present a systematic description of the holographic duals of WCFTs. It is argued that the minimal setup is not Einstein gravity but an $SL(2,R) \times U(1)$ Chern-Simons Theory, which we call Lower Spin Gravity. This point of view makes manifest the definition of boundary for these non $AdS$ geometries. This case represents the first step towards understanding a fully invariant formalism for $W_N$ field theories and their holographic duals.

\end{abstract}
\pagebreak
\setcounter{page}{1}

\tableofcontents

\pagebreak

\section{Introduction}

It is believed that holography is a very generic phenomenon that extends to cases where the geometry of space-time is not Anti de Sitter ($AdS$) \cite{Guica:2008mu,Anninos:2008fx,Anninos:2008qb,Balasubramanian:2008dm}. A related fact is that many Quantum Field Theories with no parity, no Lorentz invariance and/or with generalized symmetries are believed to be described by gravitational-like theories \cite{Afshar:2014rwa} . It is even believed that cosmological setups, like de Sitter space (dS), can be described by dual field theories \cite{Strominger:2001pn,Alishahiha:2004md,Anninos:2011af}.

The main reason behind these expectations is that the entropy of black holes is given by their area in very general conditions:
\be
S_{BH} \sim \frac{Area}{\ell^{d-1}_P} 
\ee
\noindent where $\ell_P$ is a fundamental length scale in the gravitational theory in $d+1$ spacetime dimensions. Notice that this result applies to general space-time backgrounds, including $AdS$, $dS$ and flat space. Of course our benchmark for holography remains $AdS$ and its dual Conformal Field Theory (CFT)  \cite{Maldacena:1997re,Witten:1998qj,Gubser:1998bc}. It is very likely, however, that we will not be able to understand the true nature of the holographic phenomenon until it can be extended successfully, and at a well established level, to non-$AdS$ spaces. One reason to suspect this is the case is that in $AdS$ holographic screen areas and bulk volumes scale in the same way due to the particular curvature of its metric. This fact obscures slightly the nature of holography as the distinction between volumes and areas is not that clear.

Therefore, it is necessary to extend our area of study and consider other space-times that, while non-$AdS$, can still be completely understood at the same level of precision we understand $AdS/CFT$. One important case is that of extremal \cite{Guica:2008mu} or near extremal black holes \cite{Castro:2009jf}. The near horizon geometry of these objects turns out to be completely universal. Any extremal black hole in any theory, background or dimension exhibits an $SL(2) \times U(1)$ isometry group. This means that if we could understand the holographic dual to such space times we would learn about a part of phase space of quantum gravity that seems to be fundamental. Kerr/CFT is such a proposal to study this problem \cite{Guica:2008mu}. Notice, however, that postulating the existence of a dual CFT is a bold assumption as the space-time does not posses the full $SL(2) \times SL(2)$ isometries we expect to represent the global symmetries of a two dimensional CFT. Although hidden symmetries have been discovered allowing a second $SL(2)$  \cite{Castro:2010fd} and two Virasoro symmetries were found in related setups \cite{Compere:2014bia}, it would not be surprising if CFTs did not represent the minimal dual theories responsible for this bulk physics. 

In \cite{Hofman:2011zj} a new class of two dimensional Quantum Field Theories (QFTs) was presented where only the $SL(2) \times U(1)$ factor constitutes the global symmetry group. It was shown that these theories actually possess an infinite number of conserved charges satisfying the Virasoro-Kac-Moody U(1) algebra by arguments similar to those used for CFTs \cite{joe}. These theories were named Warped Conformal Field Theories (WCFTs) \cite{Detournay:2012pc}. It turns out they have enough structure to reproduce the entropy of black holes in space-times with $SL(2) \times U(1)$ isometries\footnote{These theories can also account for the entropy of (near) extremal Kerr black holes. This case is slightly singular and it requires a separate analysis.}. These space-times are called Warped Anti de Sitter ($WAdS$) and were studied in detail in \cite{Anninos:2008fx,Anninos:2008qb}. They can appear naturally in string theory where the string sigma model can be exactly solved in certain situations \cite{Azeyanagi:2012zd}. These spaces give their name to WCFTs. 

Understanding WCFTs and their holographic dictionary has important implications for other related problems as well. Over the years, the peculiarities of $AdS_2$ space-times have been discussed by several authors \cite{Strominger:1998yg,Maldacena:1998uz,Hartman:2008dq,Sen:2011cn}. This space-time presents important differences from their higher dimensional cousins. Any finite energy excitation backreacts and destroys $AdS_2$. Related to this, it is not understood how to setup boundary conditions to get a non trivial algebra of charges. From a holographic perspective, it has been argued that Conformal Quantum Mechanics (CQM) models \cite{de Alfaro:1976je} should be dual to these geometries. An important problem, however, consists in the fact that CQM does not typically have conformally invariant vacua so they can't be obviously dual to $AdS_2$. Some of these difficulties lie at the core of the problems involved in understanding the holographic description of extremal black holes. Having a precise dual field theory that could capture the physics of the $AdS_2$ factor would be a step forward. WCFTs seem to be equipped to do this and are closer in spirit to proposals involving chiral CFTs \cite{Balasubramanian:2009bg}. $AdS_2$ is also very important from an applications perspective. It has been suggested that the physics of non-fermi liquids can be captured by holographic setups exhibiting semi-local quantum criticality \cite{Liu:2009dm,Faulkner:2009wj}. These theories can provide exotic power laws in the temperature dependence of transport coefficients, as discussed in  \cite{Hartnoll:2012rj}. There exists a large class of holographic models that presents this exotic behavior, but all of them include the physics of $AdS_2$ in one way or the other \cite{Hartnoll:2012wm}. It is, therefore, of interest to understand the field theoretic models that could account for this physics. In this ways WCFTs and related QFTs could find their place in condensed matter physics applications.

We have also mentioned cosmological setups. It turns out there also exists a connection between $dS$ and $AdS_2$ spaces. The static patch of $dS_{d+1}$ space-times is conformally related to $AdS_2 \times S^{d-1}$. This fact was exploited in \cite{Anninos:2011af} to present some evidence for a version of static patch holography. Most of the problems in  realizing this setup are the same issues observed in $AdS_2$.

We have, thus, presented a number of reasons to study WCFTs and their holographic realizations. From the point of view of field theory, it would of course be of use to have several examples of these theories that could be used as a benchmark. Sadly, there is a lack of examples in the literature motivated by QFT considerations alone. One reason for this is that until now there was no fully covariant formalism that made the symmetries of WCFTs manifest. By this, we mean there was no discussion of what are the background fields that constitute the geometry to which WCFTs couple to. Notice that this formalism is of great use for usual CFTs. In that case we can construct fully covariant actions by coupling quantum fields to background metrics. In this language, the presence of a conformal symmetry is directly connected to  Weyl symmetry acting on the background metric. By using these ingredients we can see how to construct CFTs manifestly. Even more, when no action principle description is available, background field methods allow us to calculate general properties of partition functions of CFTs. The Cardy formula \cite{Cardy:1986ie} is a concrete example of this.

WCFTs do not posses Lorentz symmetry. As such, they are not expected to couple to Riemannian geometry as CFTs do. The reason is simple. Riemannian geometry describes curved spaces that exhibit Lorentz symmetry in small enough local patches. Without Lorentz symmetry this description is not natural. We will develop in this work the necessary geometric setup so we can couple WCFTs to background fields in a way that symmetries can be realized manifestly. In particular this will prove useful to derive the equivalent of Weyl symmetry in WCFT. Similar lines of research have been explored in the literature recently to understand non-Lorentz invariant theories and their holographic duals \cite{Geracie:2014zha,Geracie:2014mta,Jensen:2014aia,Bergshoeff:2014uea,Andringa:2010it,Christensen:2013lma,Christensen:2013rfa,Hartong:2014oma,Hartong:2014pma}. In these articles, the physics of Newton-Cartan structures was studied in the holographic context. We will see that this is related to the physics of WCFTs, although not equivalent.

Armed with this geometric structure (that we call Warped Geometry), it turns out to be possible to write examples of WCFTs just by using standard QFT considerations. These theories are manifestly invariant under the infinite dimensional Virasoro-Kac-Moody U(1) algebra. Furthermore, these symmetries can be readily seen from the action of a Warped Weyl symmetry which will be described in detail. It should be pointed out that the formalism is completely invariant under general diffeomorphisms even though the geometry is not Riemannian. The difference lies in the tangent space symmetries.

This point raises another interesting connection. WCFTs posses an exotic infinite dimensional symmetry that is realized in space-time. The reason that this is possible is the assumption of locality. But up to this (important) issue, it is possible that the same geometric techniques developed here could be applied to other cases where there are also exotic infinite dimensional algebras acting on target space. $W_N$ CFTs are an example of such theories. One issue that complicates the study of these theories is that we don't have a fully democratic formalism that puts the higher spin currents in these models on equal footing with the energy-momentum tensor. An example of this consists in the fact that we use conformal dimensions to classify deformations of these $W_N$ theories, singling out the action of the spin two current on other operators. This point of view makes it confusing to study deformations by the higher spin current themselves, which are irrelevant operators under this classification\footnote{In order to understand the difficulties, see the very interesting discussion of deformations in $W_N$ theories in \cite{deBoer:2014fra}.}. It would be desirable, instead, to develop a notion of renormalization group flow for these systems that makes manifest the $W_N$ symmetry. The formalism developed in this work does the equivalent for WCFTs. As we will explain these are theories with a weight 2 and a weight 1 current. Our formalism treats these currents democratically. As such one could see WCFTs as a toy model for higher spin theories. Perhaps WCFTs should be called, lower spin theories, in analogy.

There is yet another important reason to understand the background geometry that WCFTs couple to. This is related to holography. One of the fundamental tenets of holography is that the boundary values of bulk fields determine the sources the dual QFT couples to. Thus, if one is interested in constructing a holographic bulk dual to WCFTs it is imperative that we know what this geometric variables are. Since we claim the boundary geometry is not given by Riemannian geometry, we reach the interesting conclusion that the bulk dual theory is not naturally  given by Einstein gravity. This is not completely unreasonable and similar proposals have been made before. Particularly in the case of Horava-Lifshitz gravity \cite{Horava:2009uw,Horava:2010zj}. Still, there currently exist conventional setups where Warped $AdS$ space-times appear in their space of solutions. Why should we try to build a different formalism? The reason is that none of these setups is minimal. They all contain other massive fields that are not related to or required by the symmetry structure of WCFTs. Popular examples where Warped $AdS_3$ solutions are found include Topologically Massive Gravity \cite{Anninos:2008fx} and the massive vector model \cite{Guica:2011ia}. Both cases contain extra massive fields. One can understand the situation by noticing that these models exhibit a different symmetry group  locally in space-time. The appearance of $WAdS_3$ spaces relates to the existence of symmetry breaking solutions. Therefore, by considering these theories we are making unjustified assumptions about the UV behavior of these theories. Even worse, the UV behavior of these models is far from clear and probably inconsistent. We propose, based on the boundary geometry of WCFTs to consider a bulk geometry that gives dynamics to this Warped Geometry. We will show this bulk theory is given by an $SL(2)\times U(1)$ Chern-Simons model in three dimensions. We call this theory Lower Spin Gravity. This theory is to WCFTs what three dimensional pure gravity is to CFTs in two dimensions. It represents the minimal bulk construction needed to realize the symmetry algebra holographically. Furthermore, these theories posses a much healthier UV behavior and could make sense by themselves as the Chern-Simons description of Einstein gravity might \cite{Achucarro:1987vz,Witten:1988hc,Witten:2007kt}. 

We comment on one more advantage of considering a bulk where the symmetries make manifest the symmetries of the boundary theory. Generically, when one considers non $AdS$ space-times different components of the bulk fields scale with different weights under the boundary scaling symmetry. For example, in Lifshitz solutions different components of the metric exhibit different scalings. This adds a level of confusion as it makes hard to read off boundary quantities from the bulk. In the example just mentioned the time component of the metric dominates over the space components and it is not clear anymore what the dimensionality of the boundary is. This fact complicates the formulation of holographic renormalization \cite{Ross:2011gu,Andrade:2012xy} and obscures the Weyl symmetry. We will see that in Lower Spin Gravity each geometrical field shows a well defined scaling behavior. This allows for a description of the renormalization group flow in a language consistent with the symmetries of the problem along the lines of our discussion above.

The structure of this article is the following. In section \ref{prel} we offer a brief review and a concrete definition of Warped Conformal Field Theory in two dimensions. We discuss what singles out WCFTs from the larger space of theories discussed in \cite{Hofman:2011zj} and argue that they enjoy an additional symmetry not previously discussed: a type of boost. We discuss the properties of the boost current in WCFTs and introduce the necessity of a formalism to couple WCFTs to background fields. In section \ref{wgsec} we develop the fundamental notions of Warped Geometry. We do this with especial emphasis in two dimensions but also discuss the generic $d>2$ case. We develop all notions in flat space and then extend them to generic curved spaces. It is argued that there is a natural geometric structure, called scaling structure, that plays the role of a light-cone in WCFTs. In section \ref{couplesec} we explain how to couple generic WCFTs to Warped Geometry. We explain how to construct conserved charges and obtain the Warped Weyl invariance that acts on background fields when coupled to a WCFT. Lastly, we use the formalism to write two particular examples of WCFTs in two dimensions. In section \ref{holsec} we construct the holographic bulk dual to WCFTs. We call it Lower Spin Gravity. We discuss briefly how the Virasoro-Kac-Moody algebra is coded in the bulk and make contact with previous holographic construction of Warped $AdS_3$. We finish in section \ref{conclu} with conclusions.

\section{Preliminaries: Warped Conformal Field Theory}\label{prel}

In this section we give a definition of Warped Conformal Field Theory  and a brief summary of known results. Furthermore we present a concrete characterization that singles out WCFTs from the bigger space of theories with chiral scaling considered in  \cite{Hofman:2011zj}. As it turns out, there is a type of boost symmetry that can be used to completely specify this family of theories. Lastly we argue for the necessity for completely general background field methods when analyzing these theories.

\subsection{What is a WCFT?}\label{what}

Here we review briefly the definition of Warped Conformal Field Theories given in  \cite{Hofman:2011zj} and \cite{Detournay:2012pc}. We will adjust the notation conveniently to match what follows in this article. Also, we will try to be especially precise in this definition so we can single out what makes a WCFT special.

Let us define a 2d Generalized Conformal Field Theory ($GCFT_2$) to be a unitary local Quantum Field Theory in two dimensions that posses at least three global symmetries. They are translations in both coordinates and rescalings in one of them. Let us call $x$ the scaling coordinate and $t$ the non-scaling coordinate. There is no need at this point to identify $x$ and $t$ with space and time.

These symmetries are generated by the associated conserved charges $H$, $D$ and $\bar{H}$ as:
\be\label{syms}
H : x \rightarrow x + \delta x,  \quad\quad \bar{H}: t \rightarrow t + \delta t,  \quad \quad D: x \rightarrow \lambda x .
\ee

Associated to these charges there must be conserved currents as a consequence of locality. There are given by $J^\mu$, $\bar{J}^\mu$ and $J_D^\mu$ for $H$, $\bar{H}$ and $D$ respectively.

Notice that usual 2d Conformal Field Theories ($CFT_2$) are included in this group of theories. They posses the symmetries above as well as the additional symmetry $t \rightarrow \bar{\lambda} t$. In this case we think of $x$ and $t$ as light cone variables and the theory is Lorentz invariant. This is not the case for a generic GCFT.

The commutators for the above charges are:
\be
i [ D, H] = H , \quad\quad i [ H ,\bar{H}] = 0, \quad\quad i [D, \bar{H}] =0 .
\ee
These commutators imply that we can write:
\be
J_D^\mu = x J^\mu + S_D^\mu
\ee
\noindent where $S_D^\mu$ is a local operator. Furthermore, conservation of  the $J^\mu$ and $J_D^\mu$ currents impose the following relation:
\be
J^x + \partial_x S_D^x + \partial_t S_D^t =0 .
\ee
It was argued in  \cite{Hofman:2011zj} that $S_D^x$ is a dimension 0 operator, i.e. $i [D , S_D^x] = x \, \partial_x S_D^x$. As such, its two point function with itself can only be a function of $t$. This in turn implies that $\partial_x S_D^x=0$ in correlation functions, up to contact terms, for the theory to be unitary.

The currents associated to conserved charges enjoy some ambiguities. It is possible to redefine them such that the commutation relations are still satisfied as well as the conservation equations. Given that $\partial_x S_D^x=0$ we can redefine the above to shift away $S_D^x$ and $S_D^t$ completely\footnote{See, however, below for a subtlety concerning $S_D^x$.}.
\be\label{shift}
J^x \rightarrow J^x + \partial_t S_D^t, \quad \quad J^t \rightarrow J^t - \partial_x S_D^t .
\ee

 The end result is we can set
\be
J^x=0 \quad \rightarrow \quad J^t \equiv T(x) .
\ee
This implies the existence of an infinite family of charges
\be\label{vir1}
T_\xi = \int dx \, \xi(x) T(x)
\ee
\noindent where the integral is calculated over a contour where we decide to quantize the theory. We can call this a spatial slice. It does not have to coincide with a $t= constant$ surface. As a matter of fact, as long as it is not a $x=constant$ surface, the expression above is valid. We consider this is the case when defining the initial value problem in our theory.

The charges (\ref{vir1}) were shown to form a Virasoro algebra in \cite{Hofman:2011zj}.

 It also turns out that $\bar{J}^x$ is a weight zero operator. This implies $\partial_x \bar{J^x}=0$ and $\partial_t \bar{J^t}=0$, as a consequence of the conservation equation. Thus, we can define 
 \be
 \bar{J}^x = \bar{T}(t), \quad\quad \bar{J}^t= P(x) .
 \ee
 
 We can therefore build new infinite families of conserved charges as:
 \be
 \bar{T}_{\bar{\xi}} = \int dt \, \bar{\xi(t)} \bar{T}(t), \quad \quad  P_\xi = \int dx \, \xi(x) P(x) .
\ee
It was shown also in  \cite{Hofman:2011zj} that  $\bar{T}_{\bar{\xi}}$ form another Virasoro algebra and $P_\xi$ form a U(1) Kac-Moody algebra.

At this point we can make the following statement. We said that a $CFT_2$ is a particular case of the construction above. Namely, a $CFT_2$ is a $GCFT_2$ where $P(x)=0$. Then the theory possesses  just two Virasoro algebras. Furthermore, we can assemble the $J^\mu$ and $\bar{J}^\mu$ currents in a symmetric energy-momentum tensor. This implies the theory is Lorentz invariant, in addition to the symmetries required in (\ref{syms}). So, it corresponds to a case where more structure is available. As a matter of fact, one could invert the argument. If we add the requirement of Lorentz invariance, the theory has to posses a symmetric energy-momentum tensor \cite{joe} and then $P(x)$ must vanish.

This discussion was satisfying as we have pinpointed what makes CFTs special in the bigger space of GCFTs.

What is a Warped Conformal Field Theory? It corresponds to the dual minimal case: a $WCFT_2$  is a $GCFT_2$ such that $\bar{T}(t)=0$. The question that now arises is: what is special about these theories? Is there a symmetry responsible for $\bar{T}(t)=0$?

We will answer the question in the affirmative in the next section.

\subsection{What makes a WCFT special?}\label{wcftboost}

We claim that there exists an additional symmetry we can impose such that all theories consistent with it and (\ref{syms}) are WCFTs. It is given by
\be\label{bsym}
\bar{B}: t \rightarrow t + v x .
\ee
We call this symmetry a generalized boost symmetry even though $t$ could be chosen to represent time. The commutators of $\bar{B}$ with the other charges are:
\be\label{newcom}
i [ H, \bar{B}] = - \bar{H}, \quad \quad i [ D, \bar{B}] = - \bar{B}, \quad\quad i [\bar{H} , \bar{B}] = 0 .
\ee

In a similar fashion as before, the commutators imply we can write the current associated with the $\bar{B}$ symmetry as:
\be
\bar{J}_{\bar{B}}^\mu = x \, \bar{J}^\mu + S^\mu
\ee
\noindent where $S^\mu$ is a local operator. Conservation of this current implies:
\be\label{spcurr}
\bar{J}^x+ \partial_x S^x + \partial_t S^t =0 .
\ee
The new commutators fix the scaling dimensions of the $S^\mu$ operators to be -1 for $S^x$ and 0 for $S^t$.  Unitarity then implies that $S^x=0$ following the same arguments as in  \cite{Hofman:2011zj}. Furthermore using the same type of shift as in (\ref{shift}) we can redefine $S^t$ away. Notice that in this case the constraint is stronger than the results obtained for the scaling current, as a consequence of the lower scaling dimensions. We have
\be\label{spin0}
S^\mu=0 .
\ee
The result is
\be
\bar{J}^x=0 .
\ee

This is exactly the WCFT constraint. Therefore a WCFT is a GCFT where the boost symmetry (\ref{bsym}) has been also added as a requirement. Warped Conformal Theories are not Lorentz invariant but posses instead a structure that can be used to constrain the form of conserved currents in an analogous way.

We will now make one last technical comment. In  \cite{Hofman:2011zj} it was explained that for generic GCFTs, the operator $S_D^x$ could be responsible for yet another $U(1)$ Kac-Moody symmetry. We will now show that for the minimal cases corresponding to CFTs and WCFTs this can't be the case. The way to see this is to notice that the dilatation current $J_D^\mu$ needs to be shifted to absorb the changes induced by (\ref{shift}). The corresponding shifts are:
\be
J_D^x \rightarrow J_D^x + \partial_t \left( x S_D^t - t S_D^x\right), \quad \quad J_D^t \rightarrow J_D^t - \partial_x \left( x S_D^t - t S_D^x\right) .
\ee
The result produces:
\be
J_D^x = - t \, \partial_t S_D^x, \quad \quad J_D^t = x \, J^t .
\ee
So while we see we have managed to obtain the standard result for the $J_D^t$ component, free from contamination from $S_D$, there is still a remnant from $S_D^x$ contributing to $J_D^x$. If we expect the dilatation to be part of the Virasoro family (\ref{vir1}), we must have $J_D^x=0$. It is easy to see this is the case for CFTs and WCFTs.

Let us first consider CFTs. In that case the presence of Lorentz symmetry implies the existence of a second scaling symmetry $t \rightarrow \bar{\lambda} t$. The operator $S_D^x$ is easily seen to have scaling dimension 0 under this new symmetry. As we argued in the case of $x$ scaling, this necessarily implies $\partial_t S_D^x =0$ inside correlation functions for the theory to be unitary. This means immediately that we can set $S_D^\mu=0$. So there can't be other symmetries generated by $S_D^\mu$.

In the case of a WCFT a very similar argument applies. The boost symmetry acts as $i x \partial_t$. Furthermore the commutators (\ref{newcom}) imply that $S_D^x$ can't have any boost charge. The consequence is that the two point function of $S_D^x$ needs to be annihilated by the differential operator $x \partial_t$. Once again, in a unitarity theory this implies $\partial_t S_D^x=0$. Thus, there can't be any new symmetries generated by $S_D^\mu$ in WCFTs either.

\subsection{Background field methods}

Once the symmetries of a Quantum Field Theory are understood, a very powerful tool consists in coupling such a theory to external sources. In the particular case of symmetry currents, these sources consist of background fields that define the geometry the field theory lives on. This background field method is of great importance. On the one hand, it helps us to write actions that are manifestly invariant under the symmetries we are considering. On the other, variations of the action with respect to these fields give us an alternative definition for the conserved currents of these theories.

Most importantly, we can use background fields to derive general properties of the type of theories we are interested in studying even when we don't have an explicit form of the action. This a crucial tool when one considers strongly coupled theories that might have no Lagrangian description. For example, when considering a CFT, it is known that the action is invariant under Weyl rescalings of the background metric:
\be
g_{\mu \nu} \rightarrow \lambda(x) g_{\mu \nu} .
\ee
This symmetry at the level of the background fields is completely equivalent to the full conformal symmetry that acts on the Quantum Field Theory. An important consequence of Weyl invariance is that the partition function of a generic CFT inherits this symmetry.
\be
Z[ g_{\mu \nu} ] \sim Z[\lambda(x) g_{\mu \nu}]
\ee
\noindent where the $\sim$ indicates possible quantum anomalies. Actually these anomalies are of great interest and background field methods provide one of the best ways to study them. In the case of CFTs in two dimensions these methods are directly behind the proof of the Cardy formula for the entropy:
\be
S_{CFT} = 2 \pi \sqrt{\frac{c}{6} L_0} + 2 \pi \sqrt{\frac{\bar{c}}{6}\bar{ L}_0} .
\ee

We are interested in developing this technology for the case of WCFTs. One reason is that non trivial examples of WCFT lagrangians are lacking\footnote{See, however,  \cite{Compere:2013aya,Compere:2013bya} for a proposal.}. By developing a background field method we expect to be able to construct explicit WCFTs. We will do just that in section \ref{free} and present two examples using this technology. Furthermore we expect that general results about WCFTs can be understood when an action principle is not available as we discussed above. The first question that needs to be asked is: what is the equivalent of Weyl invariance in WCFT? We will answer this question in the following section. Although we don't discuss this explicitly, this is the formalism behind the Cardy-like formula for WCFTs obtained in \cite{Detournay:2012pc}. These methods can be further exploited to perform a fully field theoretic calculation of entaglement entropy in WCFT \cite{inprog}.

It is crucial to make the following observation. Because WCFTs are not Lorentz invariant theories they do not couple naturally to Riemannian geometry. Riemannian geometry is the theory of curved spaces that locally posses Lorentz invariance. This is why we use this framework to describe background fields that couple to Lorentz invariant QFT currents. Therefore, we need to develop a different geometry that can couple to WCFTs making use of the local boost symmetry. It turns out that this setup is similar to a Newton-Cartan structure \cite{Geracie:2014zha,Geracie:2014mta,Jensen:2014aia,Bergshoeff:2014uea,Andringa:2010it,Christensen:2013lma,Christensen:2013rfa,Hartong:2014oma,Hartong:2014pma}. In general $d>2$ this will not be exactly the case, but the situation is similar. Futhermore, WCFTs posses a scale invariance. We will add a precursor to this symmetry that we will call a scaling structure. This will play in WCFT the same role that the light-cone plays in CFT and will allow us to make progress in singling out the background geometry. We call this construct Warped Geometry and we develop it in the next section.

One last comment is in order. There is another very important reason to develop background field techniques: Holography. Coupling a QFT to background geometry is the first step towards constructing a holographic dual. The way the usual holographic dictionary works is that bulk fields within dynamical geometries construct quantum field theories on the boundary of such a bulk. The boundary values of these dynamical fields constitute the backgrounds sources for the dual QFT. Therefore, in order to construct a bulk theory we need to know what kind of fields  we expect to find at the boundary. We elaborate on the systematics of this construction in section \ref{holoprop}.

Notice that because the boundary geometry will not correspond to Riemannian geometry we do not expect to find Einstein gravity in the bulk of these holographic theories. As it turns out we will find a different  theory that we call Lower Spin Gravity, which we discuss in section \ref{lowerspin}.

\section{Warped Geometry}\label{wgsec}

Let us now discuss in general terms the properties of a generalized geometry which realizes the symmetries of our WCFTs. We will call this mathematical construct Warped Conformal Geometry (WCG). This will be to WCFTs what Conformal Geometry is to CFTs.  It will be useful to leave the scaling generator aside for most of the discussion to see what can be obtained from the boost symmetry alone. Therefore we will construct WCG by first defining Warped Geometry (WG), as opposed to the usual Riemannian geometry. Then we will add some structure to this notion and build WCG. We hope this nomenclature will not induce confusions.

In 2 dimensions we will show WG to be essentially equivalent to a Newton-Cartan structure \cite{Geracie:2014zha,Geracie:2014mta,Jensen:2014aia,Bergshoeff:2014uea,Andringa:2010it,Christensen:2013lma,Christensen:2013rfa,Hartong:2014oma,Hartong:2014pma} where a scaling structure, to be defined below,  is also added. The situation will be somewhat different in $d>2$. First we will discuss $d=2$ and we will comment on $d>2$ briefly later on.

\subsection{Warped Geometry in $d=2$ flat space}

We will start by defining the symmetries that interest us in flat space and then generalize to arbitrary backgrounds.  They are (in some coordinates):

\be
x^a \rightarrow x^a + \delta^a, \quad x^a \rightarrow \Lambda^a_{\phantom{a} b}  x^b, \quad x^a \rightarrow \lambda^a_{\phantom{a} b}  x^b
\ee

\noindent where $\delta^a$ represents translational invariance, $\Lambda$ is a generalized boost transformation and $\lambda$ a rescaling of one of the coordinates. Notice that as we have stressed before we have not specified which coordinate, $t$ or $x$, should be thought as time. In many situations\footnote{The WCFT dual to (near) extremal Kerr black holes seems to be a notable exception to this.} $x$ will be thought of as space and $t$ as time. Notice that in that case, calling $\Lambda^a_{\phantom{a} b}$ a boost generator is a slight misnomer. We will use this terminology nonetheless. 

 The way we have written the equations above is meant to emphasize the similarity with Lorentz transformations. In coordinates $x^a= (x, t)$ we can write the action of the boost as:

\be\label{boost}
\Lambda^a_{\phantom{a} b} = \left( \begin{array}{cc}
1 & 0\\
v & 1 \end{array} \right), \quad \quad x \rightarrow x, \quad\quad t \rightarrow t + v x .
\ee

Scalings are

\be
\lambda^a_{\phantom{a} b} = \left( \begin{array}{cc}
\lambda & 0\\
0 & 1 \end{array} \right),  \quad \quad x \rightarrow  \lambda x,  \quad\quad t \rightarrow t .
\ee

Let us leave the scaling symmetry aside for a while and concentrate on boosts to define WG. This is nothing else than the symmetry group behind a Newton-Cartan structure in $d=2$.

The fact that we have translation invariance means we have to use vectors and not points to construct any invariant as in usual euclidean geometry. These vectors, which we call long representations for reasons that will be clear below, must transform under the remaining boost symmetry in the usual way (as coordinates do):
\be
\bar{V}^a \rightarrow \Lambda^a_{\phantom{a} b}  \bar{V}^b .
\ee

In the usual euclidean case the first non trivial tensor has two indices and represents the invariant metric. Here we notice already the existence of a boost invariant vector:

\be
\bar{q}^a = \Lambda^a_{\phantom{a} b}  \bar{q}^b \quad \rightarrow \quad \bar{q}^a = \left( \begin{array}{c} 0 \\ 1 \end{array} \right) .
\ee
 
Still we are interested in computing bilinear invariants of vectors so we can find an analogous concept to a metric. We are then looking for a symmetric tensor with lower indices such that
\be
g_{c d} = \Lambda^a_{\phantom{a} c}\,  g_{a b} \, \Lambda^b_{\phantom{a} d}   \quad \rightarrow \quad g_{c d} = \left( \begin{array}{cc} 1 & 0 \\ 0 & 0 \end{array} \right) .
\ee

It is clear that the metric is degenerate in this geometry as we can observe from:
\be
g_{a b} \bar{q}^b = 0 .
\ee

 Still we can define invariant scalar products now as:
\be
\bar{U} \cdot \bar{V} = \bar{U}^a g_{a b} \bar{V}^b = \bar{U}^x \bar{V}^x .
\ee

Notice that it is possible to define the invariant one-form $q_b=(1,0)$ as $q_b=\Lambda^a_{\phantom{a} b} q_a$. The metric satisfies

\be
g_{a b} = q_a q_b .
\ee

We can also define the antisymmetric tensor $h_{a b}$ as:
\be
q_a \equiv h_{a b} \bar{q}^b .
\ee

One can easily check that $h_{a b}$ is also invariant under the boost transformation. This two-form is non degenerate and can be inverted to $h^{a b}$ such that $h^{a b} h_{b c} = \delta^a_c$. Notice that the metric is not invertible. It is possible, however to define an upper index metric as
\be
\bar{g}^{a b} = \bar{q}^a \bar{q}^b = h^{a c} h^{b d} q_c q_d = h^{a c} h^{b d} g_{c d} .
\ee

Thus, we see that $h$ is naturally used to raise and lower indices as $\bar{V}^a = h^{ab} V_b$. The bars remind us that this is not the usual action of a metric tensor to raise indices. For example:

\be
V_a \bar{V}^a = 0 .
\ee
 
Norms of vectors can be obviously defined as $\|\bar{U}\|^2= \bar{U} \cdot \bar{U}$. It is a direct consequence of the formulae above that it is pointless to define angles in the usual way as
\be
\frac{\bar{U} \cdot \bar{V}}{\| \bar{U} \| \|\bar{V}\|} =1 .
\ee

This will be important when we consider scaling symmetries as we know that the usual conformal transformations preserve angles. 

It turns out that there is an alternative definition which will fit our purposes. Let us use the antisymmetric tensor and define a cross product ``$\times$":

\be
\bar{U} \times \bar{V} = \bar{U}^a h_{a b} \bar{V}^b = \bar{U}^x \bar{V}^t - \bar{U}^t \bar{V}^x .
\ee

This leads to the natural definition of  ``angle":
\be
\theta(\bar U,\bar V) = \frac{ \bar U \times \bar V}{\| \bar{U} \| \|\bar{V}\|}  .
\ee 

We could then parameterize vectors in ``polar" coordinates as:
\be
\bar V = \left( \begin{array}{c} \| \bar{V} \|  \\ \| \bar{V} \| \theta \end{array}\right) .
\ee 

Notice that $\theta$ transforms as $\theta + v$ under boosts as usual angles do under euclidean rotations. For future reference notice also that $\theta \rightarrow \lambda^{-1} \theta$ under rescalings. 

An interesting fact is that under these symmetries there are actually short vector representations whenever $ \| \bar{V} \| =0$. In that case it corresponds to a trivial scalar representation $\phi$ proportional to $\bar{q}^a$ as it is invariant under all symmetries (including rescalings).
\be
\| \bar{V} \| =0 \quad \rightarrow \quad \bar V^a = \phi \, \bar q^a .
\ee 

For these representations $\phi$ constitutes the invariant much as $\| \bar{V} \|$ was for the long representations. Therefore, these representations are just constructed from a scalar and the invariant vector $\bar{q}^a$.

Up until now we left out the rescaling to a large extent. One important remark is that the invariant tensors derived above transform nicely under rescalings. This is:

\be
\lambda^a_{\phantom{a} b} \bar{q}^b = \bar{q}^a, \quad \lambda^a_{\phantom{a} b} q_a = \lambda q_b, \quad \lambda^a_{\phantom{a} b}\lambda^c_{\phantom{a} d} g_{a b} = \lambda^2 g_{c d}, \quad  \lambda^a_{\phantom{a} b}\lambda^c_{\phantom{a} d} h_{a b} = \lambda h_{c d} .
\ee

Therefore we can construct invariants under the whole symmetry algebra including the rescaling. For short representations  there exists already an invariant of the whole algebra given by $\phi$. For long representations the smallest invariants need to be constructed from two such representations. In the usual case discussed in conformal geometry it would be the angle between vectors and the ratio of their lengths. In our case, because $\theta$ transforms under rescalings,  the actual invariant under $\lambda^a_{\phantom{a} b}$ is given not by the angle but by:
\be
I(\bar U, \bar V) = {\sqrt{\| \bar{U} \| \|\bar{V}\|}} \, \theta(\bar U, \bar V) .
\ee

The ratio of the norms is also an invariant, as it is for the usual rescalings:
\be
K(\bar U, \bar V) = \frac{\| \bar U \|}{\| \bar V \|} .
\ee

\subsection{Warped Geometry in $d>2$}

Let us comment briefly on how to extend the above discussion to $d>2$. There are two ways of doing this extension. One is the way taken by the usual definition of Newtonian invariance. We would add more coordinates of the type that enjoy a boost symmetry, $t$ in our case, and include a boost for each new coordinate. For reasons that will be manifest when we discuss the holographic construction of these theories, we choose to define Warped Geometry in $d$ dimensions by doing the opposite. We will extend the number of $x$ coordinates. Therefore we consider
\be
x^a = \left( \begin{array}{c} x^I \\ t \end{array}\right)
\ee

where lower case indices have the range $a=1,\ldots d$ and uppercase indices $I=1,\ldots d-1$. These geometries posses the following global symmetries:
\begin{itemize}

\item Translations: $x^a \rightarrow x^a + \delta^a$.

\item Boosts: $x^I \rightarrow x^I, \quad t \rightarrow t + v_I x^I$.

\item Dilatations: $x^I \rightarrow \lambda x^I, \quad t \rightarrow t$.

\item Euclidean rotations: $x^I \rightarrow M^I_{\phantom{I} J} x^J, \quad  t \rightarrow t$, with $M^I_{\phantom{I} J} \in SO(d-1)$. 
\end{itemize}
We will assume generically euclidean signature in $x^I$ space. These symmetries have been considered in \cite{Gibbons:2009me,Bergshoeff:2014jla} as the $z=0$ Carroll algebra.

As before, let us concentrate on Warped Geometry, leaving the rescaling out. Most concepts generalize obviously from above. There exist invariant tensors, $\bar{q}^a$ and $g_{a b}$ which satisfy the same properties as before. Notice however that the lower index tensor $q_a$ needs to be generalized to:
\be
q_a \rightarrow q_a^I .
\ee

It is not an invariant tensor anymore as it transforms under  euclidean rotations, $q_a^I \rightarrow M^I_{\phantom{I} J} q_a^J$. The metric can still be constructed as:

\be
g_{a b} = q_a^I q_b^J \eta_{I J}
\ee
where $\eta_{IJ}$ is the euclidean invariant metric.

It is interesting to note that in this case the definition of the invariant antisymmetric tensor is given by:

\be\label{vf}
q_{a_1} ^{I_1} \ldots q_{a_{d-1}} ^{I_{d-1}} \epsilon_{I_{1} \ldots I_{d-1}} = h_{a_1 \ldots a_d} \bar{q}^{a_d}
\ee
where $\epsilon$ is the totally antisymmetric euclidean invariant tensor. The existence of a totally antisymmetric tensor $h$ is of great importance as it provides us with a volume form to integrate over this space:
\be
I(\phi) = \int h_d \wedge \phi
\ee
where $\phi$ is a scalar function of the coordinates and $h_d$ is the d-form defined in (\ref{vf}). We will see in the next section that this can be generalized to curved spaces.

An important point is that there is no invariant two-form in $d$ dimensions. This means that our definition of angles between (long) vector representations as boost invariants is no longer possible. Instead, the equivalent invariant is:
\be
\theta_d(\bar U_1, \ldots \bar U_d) = \frac{h_{a_1 \ldots a_d} \bar U^{a_1}_1 \ldots \bar U^{a_d}_d}{\| \bar U_1 \| \ldots \| \bar U_d \|}
\ee
where, as before, $\| \bar U \|^2 = \bar U^a g_{a b} \bar U^b$. If we also want to construct a quantity invariant under rescalings we define, analogously to the $d=2$ case:
\be
I_d(\bar U_1, \ldots \bar U_d) =\sqrt[d]{\| \bar U_1 \| \ldots \| \bar U_d \|} \, \theta_d(\bar U_1, \ldots \bar U_d) .
\ee

\subsection{Curved Warped Geometry}

Now that we have understood the flat version of our Warped Geometry we need to extend these results to curved space. The usual recipe is to make the flat space results the geometry of a tangent space at each point in a curved manifold. We will do this in a Cartan formalism. In order to do this we need to define a map from space-time vectors in the manifold to tangent space vectors.

Let's do it first in $d=2$.

It is natural to define the warped zweibein $\tau^a_\mu$ that maps a space-time-vector in the manifold to tangent Warped Geometry variables:
\be
\tau^a_\mu: V^\mu \partial_\mu \rightarrow \bar V^a \partial_a .
\ee

In particular this constitutes a map to a long representation in tangent space. 
Notice we can construct some tangent space invariants from $\tau^a_\mu$. They are:

\be
A_\mu = q_a \tau^{a}_\mu, \quad\quad G_{\mu\nu} =   \tau^{a}_\mu \tau^{b}_\nu \, g_{a b} = A_\mu A_\nu, \quad\quad H_{\mu \nu} = \tau^{a}_\mu \tau^{b}_\nu\,  h_{a b} .
\ee

Therefore we have managed to construct the curved space analogs of $q_a$, $g_{ab}$ and $h_{ab}$. In particular we have a volume form that allows us to do integrals in curved space as:
\be
I(\phi) = \int H_2 \wedge \phi = \int d^2x \, H  \phi
\ee
where $\phi$ is a zero form, $H_2$ is the volume form and we have defined a volume density $H= \epsilon^{\mu \nu} \tau^a_\mu \tau^b_\nu h_{a b} $ using the standard Levi-Civita symbol.

Notice also that $A_\mu$, $G_{\mu \nu}$ and $H_{\mu \nu}$ are invariant under local boost transformations of $\tau^a_\mu$. This induces a gauge symmetry on the base manifold. It is therefore necessary to include the associated ``spin" connection. The full covariant derivative in the base manifold also includes the standard affine connections associated with diffeomorphism. We then define:

\be
D = \partial + \omega - \Gamma
\ee
where $\omega$ is the ``spin" connection one form and $\Gamma$ is the affine connection. The signs chosen apply to the particular case of the zweibein where this boils down to:
\be
D_\mu \tau^a_\nu = \partial_\mu \tau^a_\nu + \omega^a_{\phantom{a} b \mu}  \tau^b_\nu- \Gamma^\rho_{\mu \nu} \tau^a_\rho .
\ee

Notice that the ``spin" connection has a tensor structure completely fixed by the symmetries. Imposing that short vector representations do not transform under boosts and the specific form of the transformation matrix (\ref{boost})

\be\label{boostgen}
\Lambda^a_{\phantom{a} b}  = \delta^a_{\phantom{a} b} + v q_b \bar{q}^a  \equiv \delta^a_{\phantom{a} b} + v \bar{B}^a_{\phantom{a} b} = e^{v \bar{B}^a_{\phantom{a} b}}
\ee

\noindent we obtain 
\be
 \omega^a_{\phantom{a} b \mu} = q_b \bar{q}^a \omega_\mu
\ee
\noindent where $\omega_\mu$ transforms under a local boost as $\omega_\mu \rightarrow \omega_\mu -\partial_\mu v$. Above we defined the boost generator $\bar{B}^a_{\phantom{a} b}$.

At this point we can relate the affine connection $\Gamma$ to $\tau^a_\mu$ and $\omega_\mu$ by imposing a zweibein postulate $D_\mu \tau^a_\nu = 0$ and the invertibility  of $\tau^a_\mu$. These conditions imply:
\be\label{christ}
\Gamma^\rho_{\mu \nu} = \tau_a^\rho \partial_\mu \tau^a_\nu +  \tau_a^\rho \bar{q}^a  q_b \tau^b_\nu \omega_\mu =  \tau_a^\rho \partial_\mu \tau^a_\nu +  \bar{A}^\rho A_\nu \omega_\mu 
\ee
\noindent where we have defined $\bar{A}^\rho =  \tau_a^\rho \bar{q}^a$, which is invariant under boosts and represents the curved version of $\bar{q}^a$.

In order to characterize the geometry we introduce torsion and curvature tensors two-forms as:
\be\label{torsion1}
T^a = d \tau^a +  \omega^a_{\phantom{a} b} \wedge \tau^b  =d\tau^a + \bar{q}^a \, \omega \wedge A, \quad \quad R^a_b = d\omega^a_{\phantom{a} b } = \bar{q}^a q_b \, d \omega \equiv  \bar{q}^a q_b R .
\ee
 
In usual Riemannian geometry the next step consists in imposing the constraint $T^a=0$ in order to express the spin connection in terms of the zweibeins and therefore obtain the unique Chirstoffel connection. We note that if we do this in this case the affine connection is not fully fixed as a consequence of the ``spin" connection not being fully expressible in terms of $\tau^a_\mu$. This is a well known situation in Newton-Cartan setups. See \cite{Jensen:2014aia,Bergshoeff:2014uea} for a recent discussion.

We will argue in what follows how to proceed to deal with this ambiguity. We will require our geometry to posses some extra structure. In usual Riemannian geometry it is the metric that provides this crutch. Our Warped Geometry does not posses such a structure. It turns out that we can demand the weaker condition that our manifold is equipped with what we call a scaling structure. This nomenclature is meant to make manifest the similarity with a complex structure in standard geometry. In Lorentzian geometry it is only the information about the position of the light cone in $d=2$ that determines the complex structure\footnote{We wish to thank J. de Boer for useful remarks in this regard.}. Here we take the information to come from the scaling transformation $\lambda^a_{\phantom{a} b}$. Notice that the scaling symmetry determines two preferred axes in tangent space, as one coordinate scales and the other does not. Even before introducing the scaling symmetry, an intermediate step is to include a structure that selects these axes. We call that geometric structure a scaling structure and it is given in the $(x,t)$ coordinates by:
\be
J^a_{\phantom{a} b} = \left( \begin{array}{cc} -1 & 0\\ 0 & 0\end{array}\right) .
\ee
Conceptually, $J^a_b$ posses an eigenvalue $\Delta^a$ for each coordinate with scaling weight $\Delta^a$ defined as $x \rightarrow \lambda^{-\Delta} x$. In our case this is $\Delta=-1$ for $x$ and $\Delta=0$ for $t$. This structure separates the tangent space in two subspaces. One spanned by the eigenvector with eigenvalue equals to -1 and another spanned by the eigenvector with 0 eigenvalue. This is completely analogous to what happens in complex geometry.

More invariantly $J^a_b$ is defined as:
\be\label{scalestruct}
J^a_{\phantom{a} b} J^b_{\phantom{a} c} = - J^a_{\phantom{a} c}  \quad \textrm{such that it posses exactly one 0 eigenvalue with eigenvector}\quad \bar{q}^a .
\ee
\
We can define the eigenvector with $-1$ eigenvalue as $q^a$. Then a coordinate free expression for $J^a_{\phantom{a} b}$ is given by
\be\label{sss}
J^a_{\phantom{a} b} = -\left(q^c q_c\right)^{-1} q^a q_b .
\ee

This structure is nothing else than the generator of dilatations. As done in (\ref{boostgen}) we can express the scaling transformation for the infinitesimal transformation $\lambda= 1 + \delta\lambda$:
\be
\lambda^a_{\phantom{a} b} = \delta^a_{\phantom{a} b} - \delta\lambda J^a_{\phantom{a} b}  .
\ee

We will fix our ``spin" connection by demanding that the scaling structure is covariantly constant. This amounts to $q^a$ being covariantly constant by itself
\be\label{qconst}
0=D_\mu q^a=   \partial_\mu q^a+  \omega_\mu \bar{q}^a q_c q^c .
\ee

The above is nothing else than the requirement that the covariant derivative maps weight -1 vectors to weight -1 vectors and weight 0 vectors to weight 0 vectors. One can also think of it as requiring the geometry to have a killing vector under the boost symmetry \cite{Abbott:1982jh}.

Equation (\ref{qconst}) implies two constraints. The first one is just $q_a \partial_\mu q^a = \partial_\mu \left(q_a q^a\right)=0$. This implies we can normalize $q^a$ to satisfy $q_a q^a=1$. We can then define $\bar{q}_a$ as:
\be
\bar{q}_a q^a=0 \quad\quad \bar{q}_a \bar{q}^a=1 .
\ee

The second consequence is an expression for $\omega_\mu$:
\be\label{spinc}
\omega_\mu = -\bar{q}_a \partial_\mu q^a .
\ee

This can be seen to transform properly as the ``spin" connection should. Notice that in our $(x,t)$ coordinates this is nothing else than $\omega_\mu=0$. By performing an arbitrary local boost with parameter $v$ we see that $\omega_\mu=-\partial_\mu v$. This implies directly that the existence of a covariantly conserved scaling structure implies necessarily the vanishing curvature condition
\be
R=0
\ee
\noindent in any coordinate system. Interestingly enough, affine connections with this property are known in conventional geometry as Weitzenb\"{o}ck connections \cite{Ortin:2004ms}. 

Using $q^a$ and $\bar{q}^a$ as a preferred basis we can write:
\be
\tau^a_\mu = A_\mu q^a + \bar{A}_\mu \bar{q}^a.
\ee

The invariant vectors and space-time one-forms obviously satisfy:
\be
A^\mu A_\mu =1, \quad\quad A^\mu \bar{A}_\mu=0, \quad\quad \bar{A}^\mu A_\mu =0, \quad\quad \bar{A}^\mu \bar{A}_\mu=1 .
\ee

Using the above expression we can write the affine connection (\ref{christ}) as:
\be
\Gamma^\rho_{\mu \nu}  = \bar{A}^\rho \partial_\mu \bar{A}_\nu + A^\rho \partial_\mu A_\nu .
\ee

The torsion can also be compactly written as
\be
T^a = d \tau^a - d q^a \wedge A = q^a d A + \bar{q}^a d \bar{A} .
\ee
The torsion equations can then be written as 
\be
T = dA, \quad\quad \bar{T} =d\bar{A}
\ee
\noindent in a manifestly tangent space invariant language. This, again, looks exactly like the situation in Weitzenb\"{o}ck geometry. Notice, however that $A$ and $\bar{A}$ are fully gauge invariant under tangent space symmetries and are not vielbein. 

We should point out that an alternative point of view is to notice that we can think of $\bar{A}_\mu$ as inducing a map from the base manifold vector representations to short tangent space representations as:
\be\label{shortmap}
e^a_\mu: V^\mu \partial_\mu \rightarrow \bar{A}_\mu V^\mu \, \bar q^a \partial_a .
\ee

This $e^a_\mu$ can always be written as $e^a_\mu = \bar{A}_\mu \bar q^a$. It is clear that $\bar{A}_\mu$ transforms as a tangent space scalar. Therefore an alternative way to introduce a scaling structure is to define  two maps from manifold vectors into tangent space, $\tau^a_\mu$ and $e^a_\mu$ satisfying

\be
H_{\mu \nu} = \tau^a_{\mu} \tau^b_{\nu} h_{a b} = \tau^a_{[ \mu} e^b_{\nu] }h_{a b}= A_{[\mu} \bar{A}_{\nu]} .
\ee

This constraint is necessary as there can be at most 4 independent tangent space invariants ($A_\mu, \bar{A}_\mu$) constructed from two manifold vectors.

We are now ready to make the last step and construct the only invariants that can appear in Warped Conformal Gravity. We need to consider the scaling of the fundamental geometrical tensors $A_\mu$ and $\bar{A}_\mu$. This can be easily done by acting with the curved scaling transformation: $\lambda^\mu_{\phantom{\mu} \nu} = \lambda^a_{\phantom{a} b} \tau^b_\nu \tau^\mu_a$. By acting on $A_\mu$ and $\bar{A}_\mu$ we obtain.
\be
\bar{A}_\mu \rightarrow \bar{A}_\mu, \quad \bar{A}^\mu \rightarrow \bar{A}^\mu, \quad A_\mu \rightarrow \lambda A_\mu, \quad A^\mu \rightarrow \lambda^{-1} A^\mu .
\ee

Therefore $A_\mu$ has scaling weight $-1$ and $\bar{A}_\mu$ scaling weight 0.
From this we can see that in curved space we can build exactly one invariant out of one vector, namely $\bar{A}_\mu V^\mu$ which corresponds to what we called $\phi$ in flat space. If we have two vectors we can build the curved space generalisations of what we found in flat space. They are:
\be
I(U, V) =\frac{U^\mu H_{\mu \nu}V^\nu}{\sqrt{A_\mu U^\mu A_\nu V^\nu} }, \quad\quad K(U,V) = \frac{A_\mu U^\mu}{A_\mu V^\mu} .
\ee

The whole discussion above can be easily generalized to $d>2$. Generalizing the structure we described in flat space we note we need to consider invertible maps $\tau^{a}_\mu$ that define:
\be
A^I_\mu = \tau^{a}_\mu q_a^I, \quad \quad G_{\mu \nu}= A^I_\mu A^J_\nu \eta_{I J}, \quad \quad H_{\mu_1 \dots \mu_d} =\tau^{a_1}_{\mu_1} \dots \tau^{a_d}_{\mu_1} h_{a_1,\ldots a_d}, \quad\quad \bar{A}^\mu = \tau^\mu_a \bar{q}^a .
\ee

Notice that the invariant fields satisfy $G_{\mu \nu} \bar{A}^\mu =0$ as in flat space. So we can think of this as a curved degenerate metric where $\bar{A}^\mu$ is the degenerate vector. As before we can introduce a scaling structure $J^a_{\phantom{a} b} =  -q^{a}_I q^I_b$ with $(d-1)$ eigenvalues $-1$ and a null eigenvalue as in (\ref{scalestruct}). Equivalently we can define $e^a_\mu$ as a map to a short representation as in (\ref{shortmap}). This defines $\bar{A}_\mu$.

We can write the structure equations. Once again, there exists a frame where the ``spin" connections for each boost generator are forced to vanish in order for the scaling structure to be preserved. Then, all curvatures for them vanish trivially. The remaining Cartan equations are for the torsions and the Riemannian curvatures associated with the $SO(d-1)$ symmetry. They can readily be written in terms of $A^I_\mu$, $\bar{A}_\mu$ and $\Omega^{I}_{\phantom{I} J \mu}$, the $SO(d-1)$ spin connection, as 

\be\label{ecurv}
T^I = dA^I +\Omega^{I}_{\phantom{I} J } \wedge A^J, \quad\quad R^I_J = d\Omega^{I}_{\phantom{I} J } + \Omega^{I}_{\phantom{I} k } \wedge \Omega^{K}_{\phantom{K} J }, \quad\quad \bar{T} = d\bar{A}.
\ee

In $d$ dimensions these are the structure equations corresponding to $d$ dimensional translations and $SO(d-1)$ rotations as expected. Notice one can think of this geometry as being Riemannian (provided we demand $T^I=0$) in the $d-1$ subspace spanned by $I$ and Weitzenb\"{o}ck with respect to the symmetries connecting $I$ and the remaining non-scaling direction.

\section{Coupling a WCFT to Warped Geometry}\label{couplesec}

The developments of the last section are the necessary ingredients to couple a WCFT to a fixed background geometry. That means that if we had a concrete Lagrangian we could minimally couple it  by introducing the background fields $A_\mu$ and $\bar{A}_\mu$ or $\tau^a_\mu$ in a tangent space formulation. Also we would replace derivatives by covariant derivatives. 

Unfortunately if one lacks  an action for a WCFT in flat space, it is not possible to make progress on how to extend the theory to curved space in this naive way. Concrete examples were lacking so far in the literature, with the possible exception of the holographic construction in  \cite{Compere:2013aya,Compere:2013bya}. We will present a field theory construction of  two WCFTs in section \ref{free} and we will write the action in a fully covariant language. Most interesting theories, however, do not provide us with explicit expressions.

In the following section we will obtain some results by writing the variation of the action with respect to background fields.

\subsection{WCFT symmetries from background fields}\label{backwcft}

The generic information we can obtain comes from assuming an expression for the action in a given background $S[\tau^a_\mu, \omega_\mu]$ and studying its variations.
\be
S[\tau^a_\mu+\delta\tau^a_{\mu}, \omega_\mu+\delta\omega_\mu] = S[\tau^a_\mu, \omega_\mu] + \delta S .
\ee

What should $\delta S$ be? Our theory posses translational invariance and as such it couples to $\tau^a_\mu$ as the gauge fields of such transformations, with gauge curvature $T^a$. At the same time the theory is invariant under boosts and as such it couples to $\omega_\mu$ with trivial gauge curvature $R=0$, such that the scaling structure is preserved. Therefore we must have:
\be\label{deltaS}
\delta S = \int d^2x H \left(J^\mu_a \delta \tau^a_\mu + S^{\mu a}{\phantom{}_b} \bar{q}^b q_a \delta \omega_\mu\right)
\ee

\noindent where $H$ is the volume density $H =\epsilon^{\mu \nu} H_{\mu \nu}$, $J^\mu_a$ are the two translational currents that are naturally assembled in a generalized energy-momentum tensor and $S^{\mu a}{\phantom{}_b}$ is the boost current.  Notice $S^\mu$ plays a similar role to the usual spin current in Lorentz invariant theories.

A translational gauge transformation is defined such that $T^a$ is invariant. We see from (\ref{torsion1}) that it is given by:
\be
\delta \tau^a_\mu = -\partial_\mu \zeta^a - \bar{q}^a q_b \zeta^b \omega_\mu, \quad \delta\omega_\mu =0
\ee
\noindent where $\zeta^a$ is a tangent vector valued  gauge parameter. We have used crucially that $R=0$ in order to show that this leaves $T^a$ invariant. Demanding the invariance of the action under such transformation yields:
\be
\delta S = \int d^2x \zeta^ a \left[ \partial_\mu \left(H J^\mu_a\right) - H q_a \bar{q}^b J^\mu_b \omega_\mu \right] = \int d^2x H \zeta^ a D_\mu J^\mu_a =0 \quad \rightarrow D_\mu J^\mu_a =0 .
\ee

Notice that the variation assembles naturally in a covariant divergence that is diffeomorphic invariant and also invariant under tangent space boosts. As there is always a tangent space frame where $\omega_\mu=0$, modulo global obstructions, we can then write the familiar conservation equations for the currents:
\be
\partial_\mu \left(H J^\mu_a\right) = 0
\ee

Notice that, while this geometry does not posses a natural metric, there is still a well defined notion of an invariant vanishing divergence by the existence of the volume density $H$.

We can now play the same game for the local boost transformation. This is given by
\be
\delta \tau^a_\mu = v \, \bar{q}^a q_b \tau^b_\mu, \quad \delta \omega_\mu = - \partial_\mu v
\ee
\noindent for a gauge parameter $v$. Repeating the procedure above we obtain.
\be
\bar{q}^a q_b\partial_\mu \left(H S^{\mu b}{\phantom{}_a}\right) + \bar{q}^a J_a^\mu \tau^b_\mu q_b =0 .
\ee

This equation is completely analogous to the one that connects the antisymmetric part of the energy-momentum tensor in Lorentz invariant theories to the divergence of the spin current \cite{Ortin:2004ms}. We see that we have obtained through this background field formalism a generally covariant version of expression (\ref{spcurr}) with $S^\mu = S^{\mu a}{\phantom{}_b} \bar{q}^b q_a $. We see that the boost current is nothing else than what we called $S^\mu$ in (\ref{spcurr}).

We have been carrying the boost current $S^{\mu b}{\phantom{}_a}$ around for clarity, but as we argued in section \ref{wcftboost}, both components of $S^{\mu b}{\phantom{}_a}$ are effectively zero in WCFT. One component can be shifted away by redefining the currents while the other carries negative weight under scaling transformations which means it needs to vanish in unitary local theories. Therefore, variations of the action with respect to the spin connection $\omega_\mu$ vanish identically.

Using $S^{\mu b}{\phantom{}_a}=0$ we obtain the expression:
\be
\bar{q}^a J_a^\mu \tau^b_\mu q_b = \bar{J}^\mu A_\mu =0
\ee

\noindent where we have defined the current $\bar{J}^\mu = \bar{q}^a J_a^\mu$, which we studied in detail in section \ref{wcftboost}.

Notice we can obtain the same expression by demanding invariance of the action with respect to the space-time field variation: $\delta \bar{A}_\mu = v A_\mu$ and making use of the action variation
\be\label{actvar}
\delta S[ \delta A_\mu, \delta \bar{A}_\mu] = \int d^2x H \left(J^\mu \delta A_\mu + \bar{J}^\mu \delta \bar{A}_\mu \right) 
\ee
\noindent where we have defined $J^\mu = q^a J_a^\mu$. The expression above agrees with (\ref{deltaS}) on setting $\delta \omega_\mu=0$ from the point of view of space-time transformations. It can be seen directly that the consequences of the background field variations $\delta \bar{A}_\mu = v A_\mu$  and the tangent space boost symmetry completely agree once we set $S^{\mu b}{\phantom{}_a}=0$. 

If we also want our theory to be invariant under rescalings in such a way that the theory becomes Warped Conformal we demand that the variation of the action under $\delta A_\mu = \gamma A_\mu$ vanishes as well. The direct consequence is therefore:
\be
q^a J_a^\mu \tau_\mu^b q_b =J^\mu A_\mu =0 .
\ee

Let us finally summarize the above results for a WCFT written in flat space in tangent space coordinates such that $\omega_\mu=0$ and $A_\mu = \delta^x_\mu$, $\bar{A}_\mu = \delta^t_\mu$.
\be\label{flatcu}
J^x =0, \quad\quad \bar{J}^x =0, \quad\quad \partial_t J^t =0, \quad\quad \partial_t \bar{J}^t =0 .
\ee

These are nothing else than the basic conditions satisfied by a WCFT. These facts lead to the infinite number of conserved currents observed in \cite{Hofman:2011zj}. We shortly explain how to build conserved charges in Warped Geometry below.

\subsection{Conserved charges in Warped Geometry}

In Warped Geometry it is easy to construct conserved charges once a current is covariantly conserved.

\be
D_\mu J^\mu_{a_1 \ldots a_N} = 0 .
\ee

We have allowed a generic number of tangent space indices in the space-time currents above. Notice that it is a well known issue in non abelian theories that currents that are not gauge singlets can't be used to construct invariant charges \cite{Abbott:1982jh}. Warped Geometry however always contains the tangent space vector $\bar{q}^a$ that trivially satisfies 
\be
D_\mu \bar{q}^a =0 .
\ee

We can then always build from $J^\mu_{a_1 \ldots a_N}$ the current $J^\mu=\bar{q}^{a_1} \cdots \bar{q}^{a_N} J^\mu_{a_1 \ldots a_N}$ which is conserved in the usual sense
\be\label{conscurr}
\nabla_\mu J^\mu \equiv \frac{1}{H}\partial_\mu \left( H J^\mu\right)=0 .
\ee

Given such a current the construction of conserved quantities follows as in conventional geometry. Once we have a volume form we can define a Hodge star operator as:
\be
\star = H_{\mu \nu} .
\ee

The claim is that if $J^\mu$ satisfies (\ref{conscurr}) then 
\be
Q = \int \star J
\ee
is a conserved charge. The proof amounts to showing that the integral of $\star J$ over a closed contour $\partial V$ vanishes
\be
\oint_{\partial V} \star J = \int_V d \left(\star J\right) = \int_V H_2 \,  \frac{1}{H}\partial_\mu \left( H J^\mu\right) =0 .
\ee

In particular once we demand that our geometry preserves the scaling structure, there is another covariantly conserved tangent space vector, $q^a$. We can then define more conserved quantities for the above currents.

In the case of interest in the previous section the conserved currents are two, given by
\be
J^\mu = q^a J^\mu_a, \quad \quad \bar{J}^\mu = \bar{q}^a J^\mu_a 
\ee
and satisfy 
\be
\nabla_\mu J^\mu = 0, \quad \quad \nabla_\mu \bar{J}^\mu =0 .
\ee

In flat space coordinates we find from (\ref{flatcu}) 
\be\label{TPlab}
\left(\star J\right)_t = 0, \quad \left(\star J\right)_x \equiv T(x), \quad \left(\star \bar{J}\right)_t = 0, \quad \left(\star \bar{J}\right)_x \equiv P(x) .
\ee

This allows us to define an infinite set of the Virasoro-Kac-Moody U(1) conserved charges discussed in section \ref{what}
\be
T_\xi = \int dx \, \xi(x)T(x),  \quad\quad P_\xi = \int dx \, \xi(x) P(x) .
\ee
 
We clarify, as in section \ref{what}, that these charges make sense as long as the contours of integration chosen to quantize the theory have a non zero overlap with the $x$ direction. If we had picked $x$ as our time coordinate, the charges above would become degenerate. This is not outside  the space of possibilities.  There is no natural metric defined on this geometry. Only the background fields $A_\mu$ and $\bar{A}_\mu$. And there is no reason a priori to identify time from them.

\subsection{Warped Weyl Invariance}

The same way a CFT possess Weyl invariance when coupled to a metric $g_{\mu \nu}$, we expect our WCFTs to show a similar behavior when coupled to $A_\mu$ and $\bar{A}_\mu$. We call this invariance Warped Weyl symmetry.

These symmetries can be read directly from the variations discussed in section \ref{backwcft}. They are:
\be\label{warpedweyl}
\delta A_\mu = \gamma A_\mu \quad \quad \delta \bar{A}_\mu = v A_\mu
\ee
\noindent where $\gamma$ and $v$ are arbitrary deformation parameters that depend on space-time coordinates.

It is important to point out that we only expect these symmetries to hold classically. The change of the measure in the partition function of WCFTs will induce anomalies. These have been discussed in detail in \cite{Detournay:2012pc} and are directly responsible for the Cardy-like formula for WCFTs.

The first of the variations in (\ref{warpedweyl}) is completely analogous to $\delta g_{\mu\nu} = \lambda g_{\mu \nu}$ in CFTs. It makes explicit that only $A_\mu$ transforms under rescalings. Why do we have a second Weyl transformation? This is not a surprise. It turns out that CFTs posses  an analogous invariance given by $\delta g_{\mu \nu} = \lambda \epsilon_{\mu \nu}$. The reason this is not usually discussed is that the metric field is constrained to be symmetric. This can be done covariantly by demanding $g_{[\mu \nu]}=0$. In our case there is no obvious way to decouple the equivalent component from our background fields. Therefore we obtain a second Warped Weyl transformation.

These transformations are completely equivalent to the infinite number of conserved currents discussed in the previous section and therefore provide a powerful background method description of these systems. 

A direct consequence of this discussion is that the partition function of a WCFT defined on a non trivial background enjoys Warped Weyl symmetry, up to quantum anomalies:
\be
Z[A_\mu, \bar{A}_\mu ] \sim Z[(1+\gamma) A_\mu, \bar{A}_\mu + v A_\mu] .
\ee

Lastly we should add that our experience with $AdS/CFT$ indicates that Warped Weyl transformations will play a determinant role in the construction of holographic duals to WCFTs. In particular we expect to encounter them as symmetries of the boundary geometry. They, however, should not extend to bulk symmetries.

\subsection{The simplest WCFTs}\label{free}

So far we have been a bit abstract and discussed WCFTs in full generality. In this section we will construct the simplest examples of  WCFTs corresponding to free fermions. Although trivial as any free theory, this constitutes the first explicit example of this class of field theories showing manifestly that the class is not an empty set. Also, given the machinery developed above, we will be capable of coupling the theory to an arbitrary Warped Geometry.

As in CFT we might expect that the ``smallest" free theory is given by a fermonic theory. Furthermore, as we will see WCFTs are intimately related to chiral CFTs, as their symmetry algebras might hint. We will comment on this fact at the end of this section. Therefore, it makes sense to consider fermions as their chiral representations can be described more easily.

The first step to describe fermionic representations is to consider the gamma matrix algebra. This is described more naturally in tangent space language. Here it pays off to have developed the formalism in previous sections. 

The gamma matrix algebra will not be given by the Clifford algebra as usual but by the Warped Clifford algebra:
\be\label{wclif}
\left\{ \Gamma^a, \Gamma^b\right\} = 2 \bar{q}^a \bar{q}^b .
\ee

Notice that all we did is replace the metric in the Lorentzian case for the only available invariant symmetric two tensor our theories posses. Lower index gamma matrices are defined as we usually do in warped symmetry by:
\be
\Gamma_a = h_{a b} \Gamma^b .
\ee

This definition proves quite useful as it allows us to define our boost generator as:
\be\label{boostact}
\bar{B}= \frac{1}{8} h_{a b} \left[ \Gamma^a, \Gamma^b \right]  = \frac{1}{4} h_{a b} \left(\Gamma^a \Gamma^b - \bar{q}^a \bar{q}^b\right) = \frac{1}{4} h_{a b} \Gamma^a \Gamma^b .
\ee

One can check that it acts in the appropriate way (\ref{boostgen})  on the gamma matrices as they are in a vector representation:
\be\label{bvec}
\left[ \bar{B} , \Gamma^c \right] = \frac{1}{4} h_{a b} \left( \Gamma^a \left\{ \Gamma^b, \Gamma^c\right\} -   \Big{\{} \Gamma^c, \Gamma^a\Big\} \Gamma^b \right)=  q_a \Gamma^a \bar{q}^c .
\ee

Let us explore the consequences of (\ref{wclif}) in our usual tangent space coordinate basis $(x,t)$. It implies
\be\label{gammar}
\left( \Gamma^x \right)^2 =0, \quad\quad \left( \Gamma^t \right)^2 =1, \quad\quad \Gamma^x \Gamma^t + \Gamma^t \Gamma^x = 0 .
\ee

The road we follow is to look for the smallest non trivial representation of these expressions. As for the Lorentz case this turns out to be in terms of $2 \times 2$ matrices\footnote{The $1\times 1$ case reduces to the usual representations in terms of $\bar{q}^a$ and $q_a$ described in section \ref{wgsec}.}. Let us construct these representations.

The equation $\left( \Gamma^x \right)^2 =0$ implies that acting on a two dimensional spinor space spanned by $\Psi^0$ and $\Psi^1$
\be
\Gamma^x \Psi^0 = 0, \quad \quad \Gamma^x \Psi^1 = \gamma \Psi^0 .
\ee 

Of course we can pick the normalization to make $\gamma=1$. The only exception is the case $\gamma=0$ but that would make $\Gamma^x$ trivial and give us the trivial representation. Therefore we fix $\gamma=1$. The other restrictions in (\ref{gammar}) can be seen to give
\be
\Gamma^t \Psi^0= \beta \Psi^0, \quad\quad  \Gamma^t \Psi^1= -\beta \Psi^1, \quad\quad \beta^2=1 .
\ee
We can pick $\beta=1$ and represent the gamma matrices as:
\be
\Gamma^x = \left( \begin{array}{cc} 0 & 0 \\ 1 & 0 \end{array}\right) , \quad \quad \Gamma^t = \left( \begin{array}{cc} 1 & 0 \\ 0 & -1 \end{array}\right) .
\ee

These matrices act on spinorial representations\footnote{As for vector representations, it turns out there are short representations ($\Psi^0=0$) of these spinors that are 1 dimensional. These are trivial  as for the vectorial case.}  $\Psi^\alpha =\left(\begin{array}{c} \Psi^0 \\ \Psi^1\end{array}\right)$. In these coordinates we can write:
\be
\bar{B}= \frac{1}{4} \left(\Gamma^x \Gamma^t - \Gamma^t \Gamma^x \right) = \frac{1}{2} \Gamma^x  .
\ee
From this it is easy to see that
\be
\bar{B} \Psi^0 = 0, \quad\quad \bar{B} \Psi^1 = \frac{1}{2} \Psi^0 .
\ee

The relative factor of $\frac{1}{2}$ with respect to (\ref{bvec}) is what makes these representations spinorial.

We can also define a scaling operator by demanding that it acts on the gamma matrices in the expected way for a vector representation in this preferred basis:
\be
[ J , \Gamma^x] = - \Gamma^x, \quad\quad [J, \Gamma^t] =0 .
\ee
This is nothing else than the statement that $\partial_x$ spans the vector space with weight -1 and $\partial_t$ the vector space with weight 0.
It is easy to see that the requirements above imply
\be\label{deltapsi}
J \Psi^0 = \Delta_0 \Psi^0, \quad \quad J \Psi^1 = \Delta_1 \Psi^1, \quad \quad \Delta_0 = \Delta_1 -1 .
\ee
So,
\be
J =\left(\begin{array}{cc} \Delta_0 & 0 \\ 0 & \Delta_0+1\end{array} \right) .
\ee

Notice that the weight of each spinor is not fixed at this point. We will exploit this in a minute.

Having constructed gamma matrices and spinor representations we need to discuss how to construct other representations from them. In particular, scalar and vector representations will be necessary to construct Lagrangians. 

Let us start with scalars. In order to construct scalars we need to define the dual spinorial representation. One can easily show that the quantity  $\Psi^\alpha \epsilon_{\alpha \beta} \Psi^\beta$ is a scalar under boosts, where $\epsilon_{\alpha \beta}$ is the totally antisymmetric $2\times 2$ tensor. This prompts the following natural definition for the dual representation:
\be\label{dual}
\Psi^\dag \equiv \Psi_\alpha \equiv \Psi^\beta \epsilon_{\beta \alpha} .
\ee

We have used the symbol $\dag$ for the dual representation. This is not the standard notation when discussing Dirac fermions. We hope it will not introduce confusions. Notice that as opposed to vector representations this antisymmetric product is non vanishing as a consequence of fermionic anti-commutation properties.
\be
\Phi_- \equiv \Psi^\dag \Psi \equiv  \Psi^\alpha \epsilon_{\alpha \beta} \Psi^\beta = \Psi^0 \Psi^1 - \Psi^1 \Psi^0 = 2 \Psi^0 \Psi^1 .
\ee

Are there other scalars? For vectors we have $\bar{V}^a g_{a b} \bar{V}^b$. We suspect there is an equivalent scalar for spinors. In order to construct it we need to discuss the action of the parity operator, $P: x \rightarrow -x$. This is relevant as we will see the scalar constructed above $\Phi_-$ is actually odd under parity (as the notation suggests). We are looking to construct a parity even combination.

It turns out we already have a parity operator at our disposal. It is given by $P= \Gamma^t$. Notice that it acts as expected on definite parity operators 
\be\label{parity}
\Gamma^t \Gamma^t \Gamma^t = \Gamma^t, \quad \quad \Gamma^t \Gamma^x \Gamma^t = - \Gamma^x .
\ee

As for Lorentz spinors, there is a matrix analogous to $\gamma^5$ in our algebra, which changes sign under parity but is a scalar under boosts. This is 
\be
\Gamma^5 = q_a \Gamma^a = \Gamma^x .
\ee

One can plug $\Gamma^5$ in (\ref{boostact}) in its fully covariant form to check that it commutes with $\bar{B}$. The expression  (\ref{parity}) shows that it is parity odd.

A direct consequence of this discussion is that we can construct a parity even scalar as:
\be
\Phi_+ = \Psi^\dag \Gamma^5 \Psi = \Psi^0 \Psi^0 =0.
\ee

Unfortunately it turned out to be zero for this simple situation. We can easily fix this by considering complex spinors and considering
\be
\Phi_+ = \bar{\Psi}^\dag \Gamma^5 \Psi = \bar{\Psi}^{0} \Psi^0 
\ee
\noindent where the bar indicates complex conjugate. Because of the definition (\ref{dual}) it turns out that parity even scalars contain $\Gamma^5$ and parity odd do not as opposed to the standard Lorentzian case.

It is straightforward to construct vector representations using the gamma matrices. As above we obtain a vector $V_+$ and a pseudovector $V_-$. We consider complex spinors for generality:
\be
V^a_+ = \bar{\Psi}^\dag \Gamma^5 \Gamma^a \Psi = \left(\begin{array}{c} 0\\  \bar{\Psi}^0 \Psi^0\end{array}\right), \quad \quad V^a_- = \bar{\Psi}^\dag  \Gamma^a \Psi = \left(\begin{array}{c} \bar{\Psi}^0 \Psi^0\\ -\bar{\Psi}^0 \Psi^1 - \bar{\Psi}^1 \Psi^0\end{array}\right) .
\ee

One last comment concerns the covariant derivative acting on spinors. They do not carry space-time index so the affine connection $\Gamma^\mu_{\nu\rho}$ is not involved. There is however a spinor index. This means that the spin connection is included with a factor of the boost generator. Therefore:
\be
D_\mu \Psi = \partial_\mu \Psi + \omega_\mu \bar{B} \, \Psi = \partial_\mu \Psi  - \frac{1}{4} \left(\bar{q}_c \partial_\mu q^c\right)  h_{ab} \Gamma^a \Gamma^b \Psi
\ee
\noindent where we plugged in the curvature less spin connection (\ref{spinc}). This formula makes manifest that the spin connection does not depend on the zweibein $\tau^a_\mu$, which is not the usual situation in Riemannian geometry. In particular, in the preferred coordinate frame we have been using $\omega_\mu=0$ as we remarked before.

\subsubsection{Case I: A Warped Weyl spinor}

We now have everything we need to construct the actions associated with free theories. Let's demand that they have definite parity. We first construct a parity even action for a complex free fermion (so we can add a mass term).

\begin{eqnarray}
S_+& =& \frac{1}{4} \int dx dt  \, \epsilon^{\mu \nu} \tau^a_\mu \tau^b_\nu h_{a b} \left( i \, \tau^\rho_c \, \bar{\Psi}^\dag \,\Gamma^5 \, \Gamma^c \, D_\rho \Psi  + m \bar{\Psi}^\dag \Gamma^5 \Psi \right)\nonumber\\
& =&  \frac{1}{2} \int dx dt  \, \epsilon^{\mu \nu} A_\mu \bar{A}_\nu \left( i \,\bar{A}^\rho\, \bar{\Psi}^0 \partial_\rho \Psi^0  + m \bar{\Psi}^0 \Psi^0 \right) . \label{action+}
\end{eqnarray}

This action is manifestly invariant under both tangent space symmetries and space-time diffeomorphisms. Notice the role of the volume density $H= \epsilon^{\mu \nu} \tau^a_\mu \tau^b_\nu h_{a b}$. We have also written the action in a manifestly tangent space invariant fashion. This is the equivalent of a metric formulation in usual CFTs, where $A$ and $\bar{A}$ are the background fields. Lastly, notice that the spin connection vanishes identically in the action, as $\bar{B}^2=0$. We have thus replaced $D_\rho \rightarrow \partial_\rho$. This is in complete accordance with the vanishing of the boost current $S^\mu\sim \frac{\delta S}{\delta \omega_\mu}$.

In flat space we can substitute in $\tau^a_\mu = \delta^a_\mu$ and obtain the action
\be\label{action+f}
S_+^{flat} =\frac{1}{2} \int dx dt \,i \, \bar{\Psi}^0 \partial_t \Psi^0 + m \bar{\Psi}^0 \Psi^0 .
\ee
Notice that only $\Psi^0$ enters the action. We can think of this theory as the analog to a Weyl fermion in the Lorentzian case where only one chirality of the spinor appears. As a matter of fact this is exactly the action of a complex Weyl fermion if we allow for a mass term. The reason we don't usually add such a term is because it breaks Lorentz invariance. This is not a problem in this case. Even more, we can think of this mass term as deforming the theory and inducing a Lorentz breaking RG flow which will correspond to a WCFT as we will see.

We remarked above that the scaling weight for $\Psi^0$, $\Delta_0$ was not fixed. Here we see something interesting. If we choose it to be $\Delta_0=\frac{1}{2}$ (as for the usual Weyl fermion) the theory possess a scaling symmetry under $x \rightarrow \lambda x$. In particular once we pick this scaling the action is fully invariant under Warped Weyl symmetry (\ref{warpedweyl}). Let's see that this is the case. First we write the equations of motion.
\be
\partial_t \Psi^0 - i m \Psi^0 = 0,\quad\quad \partial_t \bar{\Psi}^0 + i m \bar{\Psi}^0 = 0 .
\ee
The solutions are given by
\be
\Psi^0 = \psi(x) e^{+ i m t}, \quad\quad \bar{\Psi}^0 = \bar{\psi}(x) e^{- i m t} .
\ee
Notice that all interesting dependence is in $x$. This is already good news as we know that in WCFT the currents are only dependent in $x$  (\ref{flatcu}). 

Now we can calculate the energy momentum tensor from the fully covariant action (\ref{action+}) by differentiating with respect to the zweibein $\tau^a_\mu$. We get:

\begin{eqnarray}
H J^\mu_a &=& \frac{\delta S_+}{\delta \tau^a_\mu} = \frac{1}{2} \epsilon^{\mu \nu}  \tau^b_\nu h_{a b} \left( i \tau^\rho_c \, \bar{\Psi}^\dag \,\Gamma^5 \, \Gamma^c \, D_\rho \Psi  + m \bar{\Psi}^\dag \Gamma^5 \Psi \right)\nonumber\\
 & & -\frac{i}{4} \tau^\mu_c  \tau_a^\rho \, \bar{\Psi}^\dag \,\Gamma^5 \, \Gamma^c \, D_\rho \Psi  \epsilon^{\sigma \nu} \tau^d_\sigma \tau^b_\nu h_{d b} .
\end{eqnarray}

In flat space this yields:
\begin{eqnarray}
J^x= J^x_x = \frac{1}{2} \left(i  \bar{\Psi}^0 \partial_t \Psi^0 + m \bar{\Psi}^0 \Psi^0 \right), & \quad \quad & J^t = J^t_x = -\frac{i}{2} \bar{\Psi}^0 \partial_x \Psi^0 , \\
\bar{J}^x =J^x_t = 0, & \quad \quad & \bar{J}^t = J^t_t = \frac{m}{2} \bar{\Psi}^0 \Psi^0 .
\end{eqnarray}

The vanishing of $\bar{J}^x$ is a direct consequence of the boost symmetry. Furthermore, we must now use the equations of motion to appreciate the full symmetries of the problem \cite{Ortin:2004ms,DiFrancesco:1997nk}. This gives us the final expressions:

\begin{eqnarray}
J^x= J^x_x = 0, & \quad \quad & J^t = J^t_x = -\frac{i}{2} \bar{\psi} \partial_x \psi , \\
\bar{J}^x =J^x_t = 0, &\quad\quad & \bar{J}^t = J^t_t =  \frac{m}{2} \bar{\psi} \psi . \label{u1cu}
\end{eqnarray}

This is a Warped Conformal Field Theory. It is clear why we considered a complex spinor. Without it there is no mass term. Notice that the mass term is responsible for making the $\bar{J}$ current non trivial. What is also important to point out is that the mass term does not spoil the scale invariance. This is possible as the $t$ coordinate does not scale in WCFT.

This example is also illuminating as far as the relation between WCFTs and chiral CFTs is concerned. As we mentioned, the action (\ref{action+f}) corresponds to the usual chiral fermion CFT deformed by a Lorentz breaking mass operator. In CFT language we would say the mass operator induces a flow to an exotic fixed point in the IR, as the mass term is relevant in CFT. On the other hand we can also take the perspective that there is no RG flow and take $m$ to be a marginal coupling under WCFT scaling. This is directly connected to the fact that the mass term can be written as a total derivative upon bosonisation. Even more, the mass $m$ can be absorbed in a rescaling of the $t$ coordinate. The only relevant information is whether $m$ is real, imaginary or zero. There is a natural interpretation of this fact from the CFT perspective. Notice from the expression of the generalized energy-momentum tensor $J^\mu_a$ that the $\bar{J}$ component is given by the U(1) current present in the chiral fermion CFT. $m$ just sets the level of the current, $\xi$. Up to charge quantization issues we can always absorb the absolute value of the level so $\xi =1,0,-1$ are the only meaningful choices. This is in complete agreement with the identification $m \sim \sqrt{\xi}$ that can be read off from (\ref{u1cu}).

From a technical point of view the only real difference between a chiral CFT with $SL(2) \times U(1)$ symmetry and a WCFT are locality issues in the target space direction associated with $U(1)$. What a WCFT does, when locality is assured, is to put on equal footing the target space symmetries of the CFT with its base space symmetries. This is of great interest in possible extensions of these techniques to other systems, like $W_N$ CFTs where a fully covariant formulation is lacking.

\subsubsection{Case II: A Warped bc system}

Let us now consider a second example that can be constructed from the spinorial representations presented above. This consists in the case of a parity odd action. It will suffice for our purposes to consider a real spinor.

\begin{eqnarray}
S_- &=& \frac{1}{4} \int dx dt  \, \epsilon^{\mu \nu} \tau^a_\mu \tau^b_\nu h_{a b} \left[ i \, \tau^\rho_c \, \Psi^\dag \, \Gamma^c \, D_\rho \Psi  + m \Psi^\dag \Psi + \frac{i}{2} D_\mu \left(  \bar{q}^a \tau^\mu_a \Psi^\dag \Psi\right) \right] \nonumber\\
 &=&  \frac{1}{2} \int dx dt  \, \epsilon^{\mu \nu} A_\mu \bar{A}_\nu \left[ i A^\rho \Psi^0 \partial_\rho \Psi^0 -     i  \,\bar{A}^\rho\, \left( 2 \Psi^1 \partial_\rho \Psi^0 \right)+ 2 m  \Psi^0 \Psi^1 \right] . \label{action-}
\end{eqnarray}

As before, the term involving the spin connection vanishes confirming our prediction $\frac{ \delta S}{\delta \omega_\mu} =0$. Here, it is a consequence of $\left(\Psi^{0}\right)^2=0$. Notice the peculiar total derivative term we added. It was chosen explicitly to make the theory exactly Warped Weyl invariant. Without it we would have obtained a generalized energy-momentum tensor with the right properties up to a possible redefinition.

In flat space $\tau^a_\mu= \delta^a_\mu$ the expression above simplifies to
\be\label{action-f}
S_+^{flat} =\frac{1}{2} \int dx dt \,\left[i \, \Psi^0 \partial_x \Psi^0 - i \left( 2 \Psi^1 \partial_t \Psi^0 \right) + 2 m \Psi^0 \Psi^1 \right] .
\ee

We see right away that the action looks scale invariant if we assign the weights $\Delta_0=0$ and $\Delta_1=1$. Luckily, this is allowed by (\ref{deltapsi}). Because the weights of the fields involved are not canonical for fermions we call this system a warped bc system, in analogy to the corresponding CFT system. Notice that the action above is invariant under Warped Weyl symmetry (\ref{warpedweyl}) if one accounts for the weights and the action of $\bar{B}$ on the spinors, i.e. $\bar{B} \Psi^1 = \frac{1}{2} \Psi^0$.

As before we calculate the equations of motion:
\be
\partial_x \Psi^0 - \partial_t \Psi^1 - i m \Psi^1 =0, \quad\quad \partial_t \Psi^0 - i m \Psi^0 =0 .
\ee
\noindent and its solutions
\begin{eqnarray} \label{sol1}
\Psi^0 = \psi^0(x) e^{i m t}, \quad \quad \Psi^1 &=& \psi^1(x) e^{- i m t} - \frac{i}{2 m} \partial_x \psi^0(x) e^{i m t}, \quad \textrm{ for} \quad m \neq 0 ,\\
\Psi^0 = \psi^0(x), \quad \quad \Psi^1 &=& \psi^1(x)  + t \,\partial_x \psi^0(x) , \quad \textrm{ for} \quad m = 0 .  \label{sol2}
\end{eqnarray}
The reader can notice that in both cases all the degrees of freedom are contained in the $x$ dependent operators $\psi^0(x)$ and $\psi^1(x)$. This suggest that the theory is a WCFT. Let us confirm this fact by calculating the generalized energy-momentum tensor:

\begin{eqnarray}
H J^\mu_a = \frac{\delta S_+}{\delta \tau^a_\mu} &=& \frac{1}{2} \epsilon^{\mu \nu}  \tau^b_\nu h_{a b} \left[ i \, \tau^\rho_c \, \Psi^\dag \, \Gamma^c \, D_\rho \Psi  + m \Psi^\dag \Psi  + \frac{i}{2} D_\mu \left(  \bar{q}^a \tau^\mu_a \Psi^\dag \Psi\right) \right] \nonumber \\ & & -\frac{i}{4} \epsilon^{\sigma \nu} \tau^d_\sigma \tau^b_\nu h_{d b} \left[\tau^\mu_c  \tau_a^\rho \, \Psi^\dag  \, \Gamma^c \, D_\rho \Psi + \frac{1}{2}\bar{q}^c \tau^\mu_c \tau^\rho_a D_\rho \left( \Psi^\dag \Psi\right)\right] .
\end{eqnarray}
Evaluating in flat space we obtain:
\begin{eqnarray}
J^x= J^x_x =  - i \Psi^1 \partial_t \Psi^0 +  m \Psi^0 \Psi^1, &\quad\quad & J^t = J^t_x = i \Psi^1 \partial_x \Psi^0 , \\
\bar{J}^x =J^x_t = - \frac{i}{2} \Psi^0 \partial_t \Psi^0, &\quad\quad & \bar{J}^t = J^t_t = \frac{1}{2} \left( i \Psi^0 \partial_x \Psi^0 + 2 m \Psi^0 \Psi^1 \right) .
\end{eqnarray}
Using the equations of motion we get the satisfying result for $m=0$:
\begin{eqnarray}
J^x= J^x_x = 0, &\quad\quad & J^t = J^t_x = i \psi^1 \partial_x \psi^0 , \\
\bar{J}^x=J^x_t = 0, &\quad\quad & \bar{J}^t = J^t_t = \frac{1}{2} \psi^0 \partial_x \psi^0 .
\end{eqnarray}
and $m \neq 0$:
\begin{eqnarray}
J^x= J^x_x = 0, &\quad\quad & J^t = J^t_x = i \psi^1 \partial_x \psi^0 , \\
\bar{J}^x =J^x_t = 0, &\quad\quad & \bar{J}^t = J^t_t = m \psi^0 \psi^1 .
\end{eqnarray}

This last result agrees with the one obtained for the Weyl spinor if we give up the idea that $\bar{\psi}$ and $\psi$ are complex conjugates and we identify $2 \psi^1 \rightarrow -\bar{\psi}$ and $\psi^0 \rightarrow \psi$. In that case we could have assigned different weights to $\bar{\Psi}$ and $\Psi$ and our results can agree.

It is satisfying that we have now concrete examples of WCFTs. This means that this class of theories can be actually as large and rich as the space of CFTs. While we have shown a clear connection between WCFTs and chiral CFTs it is not obvious if this relation holds over the whole space of theories. When it does, it is quite remarkable that the technology developed here allows us to put target space symmetries and base space symmetries on equal footing. From a purely algebraic point of view, this distinction is purely artificial. It is a direct consequence of making base space locality manifest. When target space locality exists there is no longer a reason to generate a divide. In cases where these symmetries are intertwined in a complicated way, as in $W_N$ theories, this non democratic attitude obscures some of the symmetries and complicates the study of the theory. 

One important area where one can make progress with this new formalism is in the study of the renormalization group flow of theories without Lorentz symmetry and/or exotic internal symmetries. Deformations that are complicated or might even look irrelevant from the point of view of CFT could become marginal when viewed under the right scaling operator, shedding light on the infrared behavior of such theories. An example where these ideas could have an impact is in the study of deformations by higher spin operators in $W_N$ CFTs  \cite{deBoer:2014fra}. As these operators are connected to the energy-momentum tensor by $W_N$ symmetry, it is possible that a generalized RG treatment exist where they can be viewed as marginal.

Given the importance of understanding the RG in these theories we now turn to an important tool in this regards: the construction of holographic duals to WCFTs. We will show in the next section how these duals can be constructed and how they connect with the existing literature.

\section{Building a Holographic dual for WCFTs}\label{holsec}

In this section we build a bulk holographic dual to WCFTs in a systematic fashion. Of course, there exist holographic setups where Warped $AdS_3$ solutions are present among other vacua \cite{Anninos:2008fx,Anninos:2008qb}. This is actually where WCFTs get their name. Also, WCFTs have been proposed as duals to Extremal Kerr Black Holes \cite{Detournay:2012pc}. So it seems there exist already examples in the literature.

All these examples carry important shortcomings. One is that they are described by metric theories. As such these descriptions posses a different, typically much larger, group of symmetries in the UV. The situation is akin to describing a low energy theory that does not posses a symmetry by embedding it in a UV theory with that symmetry. While the situation can be useful sometimes, if there is a physical reason (e.g. evidence for the hidden symmetry, knowledge of the UV theory, etc.) or a technical reason (e.g. the UV theory is renormalizable) it is not generic. The minimal description of our physics should not invoke unwanted components.

Related to the point above, all known holographic setups of this kind involve more bulk fields than the symmetry dictates. Two popular ways to achieve Warped $AdS_3$ space-times suffer from this condition. One is the massive vector model \cite{Taylor:2008tg}. This system possess  massive vector modes that correspond to vector operators that are not conserved. Another well studied setup corresponds to Topologically Massive Gravity   \cite{Deser:1981wh,Deser:1982vy}. This system produces Warped $AdS_3$ solutions by including higher derivative corrections to the gravity action. These terms introduce massive modes and jeopardize the UV behavior of the theory. 

Our objective for this section is to build the minimal bulk description that contains only information about the symmetries of our boundary theory without any unjustified extra ingredient. We are searching for the equivalent of Einstein Gravity in AdS space-times to standard CFTs. In general we can't expect these models to be consistent by themselves. Usually they are UV completed by e.g. string theory. The point is that at this stage we don't want to assume anything about this UV completion. Furthermore, we will pay particular attention to the case of a 3 dimensional bulk. As we will see we might have reasons to believe this theory is UV complete for similar reasons we have to suspect 3 dimensional Gravity \cite{Witten:2007kt} (or at least its Chern-Simons description \cite{Witten:1988hc}) might make sense.

\subsection{General Holographic proposal}\label{holoprop}

We present the following proposal to build a holographic dual to a boundary Quantum Field Theory with an arbitrary symmetry group that presents a scaling invariance.

First, write down the minimal set of global symmetries of the theory and its associated currents leaving the scaling symmetry out. By minimal set of symmetries we mean that we should not include symmetries that are imposed on us necessarily given the minimal set and the assumption of scale invariance. In the particular case of CFTs this means we should only include translations and Lorentz rotations. The conformal generators come for free once scale invariance is added to the mix, at least in two dimensions  \cite{joe}. 

Next, couple these currents to background fields. These fields determine the geometry our QFT couples to. We now must develop a theory of this geometry for arbitrary background fields and write all gauge covariant generalized curvature tensors for our background fields. This is nothing else than Riemannian geometry for CFTs and the Warped Geometry developed in section \ref{wgsec} for WCFTs. The endpoint of this process is given by equations of a form like (\ref{torsion1}) or (\ref{ecurv}).

The following step consists in generalizing this geometry to one space-time dimension higher. This process might not be unique. Then, impose equations of motion for the curvatures in this higher dimensional geometry. They should be such that in a minimal setting they can be built from the higher dimensional background fields (that have now become dynamical) themselves. If we have been successful the number of different equations of motion we can write matches precisely the number of parameters our QFT possess at the level of symmetries alone. For example, a parity invariant CFT is characterised at the level of the symmetry algebra by one parameter alone: the central charge $c$. The holographic dual of this situation is that under the same conditions the only free parameter for the Einstein equations of motion at leading order in derivatives is the cosmological constant in Planck units.

This discussion is a bit abstract. We will follow the steps above precisely in our case of interest and will construct a minimal holographic dual to WCFTs. We will focus mainly on a 3 dimensional bulk. Whether it is possible to carry out this program successfully in other situations for QFTs with different symmetry groups is not obvious.

\subsection{Lower Spin Gravity}\label{lowerspin}

Let us carry out the procedure outlined above. Actually we are almost done. From the discussion in section \ref{wgsec} we know that the $d$ dimensional curvatures that exist in Warped Geometry are given by (\ref{ecurv}), which we reproduce below

\be\label{ecurv2}
T^I = dA^I +\Omega^{I}_{\phantom{I} J } \wedge A^J =0, \quad\quad R^I_J = d\Omega^{I}_{\phantom{I} J } + \Omega^{I}_{\phantom{I} k } \wedge \Omega^{K}_{\phantom{K} J }, \quad\quad \bar{T} = d\bar{A}
\ee
\noindent where we remind the reader that the index $I=1 \ldots d-1$ captures only the euclidean symmetry group $SO(d-1)$ acting on the $x^I$ coordinates. Notice we set the torsions $T^I$ to 0. We do this so we can determine the spin connection $\Omega^I_{\phantom{I} J}$ from the $A^I$s as in usual Riemannian geometry. Warped Geometry is however not the same as Riemannian geometry as $I$ does not run over all coordinates ($t$ is left out). Also the associated curvatures to all boost generators have been set to zero already and we omit them in (\ref{ecurv2}). This was done to preserve the scaling structure. Notice that while we won't introduce a scale symmetry in the bulk, this requirement is critical to have an extension of the scaling structure in the bulk. One can think of this as what the extension of the light cone structure into bulk represents in usual $AdS$ holography.

One last comment about this geometry. It is possible to think of it as a dual version of Newton-Cartan. While there it is the number of $t$ variables that has been generalized, we emphasize that in the case at hand we are extending the number of scaling coordinates $x^I$. The reason behind this is grounded in holography. We expect the holographic direction to capture information about the RG flow of the theory. Therefore, it is natural to extend the scaling coordinates into the bulk. If the scaling structure had a more complicated set of eigenvalues, it is an interesting problem to figure out how to arrive at the natural bulk extension. This is a necessary point that needs to be confronted if one is interested in $W_N$ CFTs.

The choice of equations of motion is therefore dictated by what can be written on the right hand side of $R^I_J$ and $\bar{T}$ in terms of the fields $A^I$ and $\bar{A}$ themselves as $\Omega^I_{\phantom{I} J}$ is a dependent quantity. We are then looking for the most general covariant set of equations that can be built from this ingredients to leading order in derivatives (i.e. without including multiple powers of the curvature tensors) and with $T^I=0$. The answer is very simple and quoted below for the general $d$ dimensional case:

\be\label{eom0}
T^I = 0, \quad\quad R^{I J}  + c  \, A^I \wedge A^J =0, \quad \quad \bar{T} = 0 .
\ee

That is it. There are no other invariants that can be included at this order. $c$ is nothing else than a cosmological constant term and we expect it to determine the central charge of the theory as in $AdS/CFT$. Therefore we fix $c$ to be positive.

In $d=3$ the situation is just slightly different. In that case there exists an $SO(2)$ antisymmetric tensor $\epsilon_{I J}$ and we can write:
\be\label{eom1}
T^I = 0, \quad\quad R^{I J} + c  \, A^I \wedge A^J + a \, \bar{T} \epsilon^{I J}=0, \quad \quad \bar{T} + b  A^I \wedge A^J \epsilon_{I J} + d \,R^{IJ}\epsilon_{I J}=0 .
\ee

Some of these extra terms can be dealt with by diagonalising the equations of motion above as:
\be\label{eom2}
T^I = 0, \quad \quad R^{I J} +\frac{c - 2 ab}{1 - 2 a d} \, A^I \wedge A^J=0  \quad \quad \bar{T} + \frac{b - dc}{1- 2 a d}  A^I \wedge A^J \epsilon_{I J} =0 .
\ee
With the exception of the degenerate case $2 ad =1$ (which would violate the assumption of vielbein invertibility), we can then always set $a$ and $d$ to zero. We are left with the following equations of motion in components:
\begin{eqnarray}
d A^1 - \Omega \wedge A^2 = 0, \quad \quad  d A^2  +\Omega \wedge A^1 = 0,  \\
d \Omega - c \,A^1 \wedge A^2 = 0, \quad \quad d\bar{A} + 2 b\, A^1 \wedge A^2 = 0 ,
\end{eqnarray}
\noindent where we have defined $\Omega \equiv \Omega^2_{\phantom{2} 1}$.
These equations of motion can be derived from the following Chern-Simons action.
\be
S = \kappa \int_{bulk} A^1 \wedge d A^1 + A^2 \wedge dA^2 + 2 \, \Omega \wedge A^1 \wedge A^2 - \frac{  \alpha \, b^2 +1}{  c} \Omega \wedge d\Omega - \frac{  \alpha \, c}{ 4} \bar{A} \wedge d \bar{A} - \alpha \, b \, \Omega \wedge d \bar{A}
\ee
\noindent where $\kappa$ is an overall normalization and $\alpha$ is a free parameter. Notice that $|\alpha|$ can be absorbed from the action by $\bar{A} \rightarrow \frac{\bar{A}}{\sqrt{|\alpha|}}$ and $b \rightarrow  \frac{b}{\sqrt{|\alpha|}}$. Therefore only the sign of $\alpha$ has physical meaning.  In order to see what this action corresponds to we can make the following field redefinition.

\be\label{ident}
 \bar{A}  \equiv \sqrt{\left|\frac{4}{\kappa c \alpha}\right|} \, \bar{B} - \frac{2 b}{c} B^3, \quad\quad A^1 \equiv \frac{B^1}{\sqrt{c}}, \quad\quad A^2 \equiv \frac{B^2}{\sqrt{c}}, \quad\quad \Omega  \equiv B^3 .
\ee

The action becomes that of the $SL(2) \times U(1)$ Chern-Simons model

\be\label{csa}
S = \frac{k}{2}\int_{bulk} B^1 \wedge d B^1 + B^2 \wedge dB^2 - B^3 \wedge dB^3+ 2 \, B^1 \wedge B^2 \wedge B^3 - \xi \int_{bulk} \bar{B} \wedge d \bar{B}
\ee
\noindent where $k = \frac{2 \kappa}{c}$ and $\xi = +1, 0 -1$ is just the sign of $\alpha$. As in the dual WCFT description we end up with one free continuous parameter, $k$, that will determine the central charge and one discrete parameter, $\xi$, that decides the sign of the level of the $U(1)$ Kac-Moody algebra. Notice that $b$ and $|\alpha|$ only appear in connecting $SL(2) \times U(1)$ variables to Warped Geometric variables and are not fundamental parameters of the theory.

This is the gravitational dual to a WCFT. In analogy to the discussions of $SL(N)$ Chern-Simons theory, we call this theory Lower Spin Gravity. Notice that this is exactly what this theory is. It describes the dynamics of metric field $G_{\mu \nu} = A^I_\mu A^J_\nu \eta_{I J}$ satisfying a chirality condition $G_{\mu \nu} \bar{A}^\nu=0$ and a (invertible) gauge field $\bar{A}_\mu$ in a completely democratic way, same way as $SL(N)$ does for higher spin fields. One should notice that we are abusing a bit the notation by calling $G_{\mu \nu}$ spin 2 and $\bar{A}_\mu$ spin 1 as they are really classified by the boost symmetry operator $B$ in our geometry and their weight.

In the following subsections we will re-obtain the Virasoro-Kac-Moody U(1) algebra we expect the bulk to exhibit as non trivial gauge transformations in a way completely analogous to the well known $AdS_3$ case. Finally we make connection between this theory and metric descriptions of Warped $AdS_3$.

\subsection{Boundary Conditions and asymptotic symmetries}

In this section we define boundary conditions for our Lower Spin Gravity. We will do so at the level of our $SL(2) \times U(1)$ Chern-Simons action and then translate what this means for our geometric variables given by the $A$s and $\bar{A}$. This problem has been studied in detail in the literature. We will  adapt the results of \cite{deBoer:2014fra,Campoleoni:2010zq,deBoer:2013gz} to our case of interest. The discussion follows closely the references above.

Let us introduce the standard Chern-Simons notation for the connection in terms of Lie Algebra valued connections, $B = B^\ell J_\ell$ where $J_\ell$ are $SL(2)$ generators. Let us consider the basis of $SL(2)$ given by, $L_+$, $L_-$ and $L_0$
\be
[ L_+, L_-] = 2 L_0, \quad\quad [L_0, L_+] = - L_+, \quad\quad [L_0, L_-] = L_- .
\ee

Writing $B$ in components we can relate this basis with the one used in writing the action (\ref{csa}).
\be
B^1 = B^+ - B^-, \quad\quad B^2 = B^0, \quad\quad B^3= B^+ + B^- .
\ee

We can choose to normalize the generators by picking their trace as $Tr[ L_0^2]=\frac{1}{2}$ and $Tr[L_+ L_-] =-1$. The Chern-Simons action becomes
\be
S = k \int_{bulk} Tr\left[ B \wedge d B + \frac{2}{3} \, B \wedge B \wedge B\right] - \xi \int_{bulk} \bar{B} \wedge d \bar{B}
\ee
\noindent and the equations of motion become the condition that both $B$ and $\bar{B}$ are respectively $SL(2)$ and $U(1)$ flat connections. In finding solutions to these constraints one can always decide to make a gauge choice. We write the connections as
\be\label{gaugesol}
B = \beta^{-1} d \beta + \beta^{-1}\left(L_+ dx + \gamma \right) \beta, \quad\quad \bar{B} = dt + \bar{\gamma}
\ee
\noindent where, if we parameterize our three dimensional bulk by coordinates $(x,t, \rho)$,  $\beta=e^{\rho L_0}$ is an element of the $SL(2)$ algebra and $\gamma(x,t)$ is a $\rho$ independent connection. Notice the constant shifts introduced in defining $\gamma$ and $\bar{\gamma}$. This will be convenient when setting boundary conditions. Explicitly, we have
\be\label{conec}
B = L_0 d\rho + e^{\rho} L_+ d x + e^{-\rho L_0}  \, \gamma \, e^{\rho L_0}, \quad \quad \bar{B} = dt + \bar{\gamma} .
\ee

We can use this notation to write the variation of the action. On-shell this reduces, as usual, to a boundary term:
\begin{eqnarray}
& &\delta S = -k \int_{bdy}  Tr\left[ \gamma \wedge \delta \gamma + L_+ dx \wedge \delta \gamma \right] + \xi \int_{bdy} \bar{\gamma} \wedge \delta \bar{\gamma} + dt \wedge \delta \bar{\gamma}\\
 &=&- k \int_{bdy} dx dt \, Tr\left[ \gamma_x  \delta \gamma_t  -  \gamma_t \delta \gamma_x + L_+ \delta\gamma_t \right] +\xi \int_{bdy} dx dt \,\, \bar{\gamma}_x \delta \bar{\gamma}_t -  \bar{\gamma}_t  \delta \bar{\gamma}_x - \delta\bar{\gamma}_x . \nonumber
\end{eqnarray}

The curious looking terms that are linear in $\gamma$ and $\bar{\gamma}$ play the same role as the shift of the connections by some constant $SL(2)$ generator in  \cite{deBoer:2014fra,deBoer:2013gz}. The fact that the $U(1)$ current gets such a term is connected to the fact that it has acquired a geometric role in our construction.

A comment is in order about the boundary geometry. We pick the boundary to posses a flat Warped Geometry. This is natural as the boundary theory will be a WCFT. This means that the volume form is just unity, as used above. Importantly for what follows, there exists a scaling structure in this geometry. This plays the role of a complex structure in usual $AdS/CFT$. This way we will get away without ever defining a metric on the boundary. In order for our variational problem to be well defined we add a boundary term of a similar type to the one discussed in \cite{deBoer:2014fra,Campoleoni:2010zq,deBoer:2013gz} :
\be\label{bdyterm}
S_{bdy} = - k \int_{bdy} dx dt \, q_a \epsilon^{a b} \bar{q}_b  \, q^c \bar{q}^d \,Tr\left[\gamma_c  \gamma_d \right] + \xi \int_{bdy} dx dt \, q_a \epsilon^{a b} \bar{q}_b  \, q^c \bar{q}^d \, \bar{\gamma}_c \bar{\gamma}_d + \xi \int_{bdy} dx dt \, q_a \epsilon^{a b} \bar{q}_b  \, q^c \,  \bar{\gamma}_c .
\ee
As we mentioned, we managed to write the integrals above with the help of the scaling structure $q^a \bar{q}^b$ (\ref{sss}). This selects the preferred axis $t$ and $x$, as the complex structure does for the light-cone in the usual Lorentz invariant case. We stress there is no metric on the boundary.

The full action has the following variation:
\begin{eqnarray}
\delta S_{full} =&-& 2 k \int_{bdy} dx dt \, q_a \epsilon^{a b} \bar{q}_b  \, q^c \bar{q}^d Tr\left[ \gamma_c  \delta \gamma_d  \right] -k \int_{bdy} dx dt \, q_a \epsilon^{a b} \bar{q}_b  \,  \bar{q}^c Tr\left[ L_+ \delta \gamma_c  \right]  \nonumber \\ 
&+&2 \xi \int_{bdy} dx dt \, \, q_a \epsilon^{a b} \bar{q}_b  \, q^c \bar{q}^d  \bar{\gamma}_c \delta \bar{\gamma}_d .
\end{eqnarray}

We see the variational problem is well defined if we fix $\bar{q}^c \gamma_c=0$ and $\bar{q}^c \bar{\gamma}_c=0$. In $(x,t)$ coordinates the variation reads:

\begin{eqnarray}
\delta S_{full} =&-& 2 k \int_{bdy} dx dt \,Tr\left[ \gamma_x  \delta \gamma_t  \right] -k \int_{bdy} dx dt \,Tr\left[ L_+ \delta \gamma_t  \right]  \nonumber  \\ 
&+&2 \xi \int_{bdy} dx dt \, \,  \bar{\gamma}_x \delta \bar{\gamma}_t 
\end{eqnarray}
\noindent and all we are doing is fixing $\gamma_t$ and $\bar{\gamma}_t$ at the boundary.

We consider solutions of a similar form to the ones used in $AdS_3$ \cite{Campoleoni:2010zq}:
\be\label{solus}
\gamma =\left(\frac{T(x)}{k} - \frac{P(x)^2}{k \xi}\right) L_-dx, \quad\quad \bar{\gamma }=  \frac{P(x)}{\xi} dx .
\ee

These asymptotics  are chosen such that the boundary warped geometry is constructed from the leading pieces that were removed explicitly from the expressions above in (\ref{conec}). For any arbitrary functions $T(x)$ and $P(x)$ these are flat connections. Given these solutions one can calculate the variation of the on-shell action with respect to the boundary warped geometric data $q^a$ and $\bar{q}^a$. If our theory yields a holographic description of the boundary WCFT we expect this calculation to give the currents $J^a$ and $\bar{J^a}$ in (\ref{actvar}). Part of the variation comes from changing this data directly. Only (\ref{bdyterm}) is affected by such change as the bulk action is topological. There is a second contribution. Notice that when calculating this variation the boundary conditions $\bar{q}^a \gamma_a=\bar{q}^a \bar \gamma_a=0$ need to be maintained as to not change the Dirichlet problem. This implies that the boundary values of the fields need to change as
\be
\bar{q}^a \delta \gamma_a = q^a \gamma_a \bar{q}^b \delta q_b + \bar{q}^a \gamma_a \bar{q}^b \delta \bar{q}_b, \quad \bar{q}^a \delta \bar\gamma_a = q^a \bar\gamma_a \bar{q}^b \delta q_b + \bar{q}^a \bar\gamma_a \bar{q}^b \delta \bar{q}_b .
\ee

Including these effects the total action variation on shell is

\be
\delta S_{full} =\int dx dt \, T(x) \delta q_t + P(x) \delta\bar{q}_t .
\ee

This matches exactly the field theory result (\ref{actvar}) in flat space. We have thus managed to calculate the conserved currents of our WCFT holographically and can identify $T(x)$ and $P(x)$ with the WCFT values (\ref{TPlab}).

We can calculate the algebra of these currents by repeating the calculation in \cite{Campoleoni:2010zq}. One can ask what is the set of gauge transformations that leave the form of the solutions (\ref{gaugesol}) invariant. That is we are looking for $\varphi$ and $\bar{\varphi}$ such that
\be
\delta \left(L_+ dx + \gamma \right) = d\varphi + [\left(L_+ dx + \gamma \right), \varphi]  = \left(\frac{\delta T}{ k} - 2 \frac{P \delta P}{k\xi}\right) dx \,  L_-   \quad \delta \bar \gamma= d \bar{\varphi} = \frac{ \delta P}{ \xi} dx
\ee
does not change the form (\ref{solus}). It turns out the most general solution is parameterised by functions $\epsilon(x)$ and $\bar{\epsilon}(x)$ as
\be
\varphi = \epsilon L_+  - \partial \epsilon \, L_0 + \left(\frac{1}{2} \partial^2 \epsilon + \left(\frac{T}{k} -\frac{P^2}{k \xi}\right) \epsilon \right) L_-, \quad\quad \bar{\varphi}=\frac{\bar{\epsilon}}{2}+ \frac{\epsilon\, P}{\xi} .
\ee
Under these transformations, $T(x)$ and $P(x)$ transform as:
\be
\delta T =  \epsilon \partial T + 2 T \partial \epsilon+ \frac{k}{2} \partial^3 \epsilon + P \partial \bar{\epsilon} \quad\quad \delta P =  \frac{\xi}{2} \partial \bar{\epsilon} +\epsilon \partial P + P \partial \epsilon  .
\ee

These are the same Schwarzian derivatives presented in \cite{Detournay:2012pc}\footnote{There, different sign conventions were used.} for the Warped Conformal algebra. We can read the Virasoro central charge $\hat{c}_{vir}$ and the $U(1)$ Kac-Moody level $\hat{k}_{u(1)}$ from these expressions:
\be
\hat{c}_{vir} = 6 k, \quad \quad \hat{k}_{u(1)} = \xi.
\ee

We conclude the bulk posses the same infinite dimensional family of conserved symmetry charges as a WCFT. It is a straightforward exercise to use the algebra above to compute charges as in \cite{Campoleoni:2010zq}.

\subsection{Warped $AdS_3$ space-times from Lower Spin Gravity }

We can now use the form of $\gamma$ and $\bar{\gamma}$ to write down $B$ and $\bar{B}$ explicitly using (\ref{conec}) as

\be\label{bfields}
B^0 = d \rho, \quad\quad B^+ = e^\rho dx, \quad\quad B^- = e^{-\rho} \left(\frac{T(x)}{k} - \frac{P(x)^2}{k \xi}\right) dx, \quad\quad \bar{B}= dt + \frac{P(x)}{\xi} dx .
\ee

We see explicitly that each component is invariant under rescalings $\rho \rightarrow \rho - \log \lambda$, $x \rightarrow \lambda x$ if we assign weight 1 to $P(x)$ and weight 2 to $T(x)$ under the transformation. This is the statement that the dual theory has a scale invariance. Furthermore we see that the Warped Weyl symmetry is explicit in this form. The expressions above retain the same form under the changes of coordinates responsible for Warped symmetry \cite{Detournay:2012pc}.
\be
x \rightarrow f(x), \quad\quad t \rightarrow t + g(x) .
\ee

Warped Weyl invariance (\ref{warpedweyl}) with generic coordinate dependence can be seen by allowing more general bulk diffeomorphisms.

These fields posses a definite scaling weight. One could now go back to (\ref{ident}) to express the original $A$ and $\bar{A}$ fields in terms of the expressions above. Notice that the extra parameters  involved are not fully physical, i.e. they do not appear in (\ref{csa}). They correspond to a particular identification of geometry variables $A, \bar{A}$ from the fundamental Chern-Simons variables $B, \bar{B}$. They are analogous to the $AdS$ radius in usual $AdS_3 /CFT_2$ which is not a physical variable, only its value in planck units is.

As this identification of the $A$s is a bit arbitrary, in what follows we will consider a generalization of  (\ref{ident}). Let us take the $SL(2)\times U(1)$ Chern-Simons action (\ref{csa}) as the fundamental theory and consider defining $A^1$, $A^2$, $\Omega$ and $\bar{A}$ from a generalized identification. Define three linearly independent vectors in $SL(2)$,  $(\zeta_0^\ell, \zeta_1^\ell, \zeta_2^\ell)$ and the inverse vectors $(\hat{\zeta}^0_\ell, \hat{\zeta}^1_\ell, \hat{\zeta}^2_\ell)$  such that
\be
\hat{\zeta}^I_\ell \, \zeta_J^\ell = \delta^I_J \quad \quad \textrm{for} \quad I,J = 0,1,2 .
\ee

These vectors define a particular identification of Chern-Simon fields $B$ with the geometric variables $A$ as:

\be
 \bar{A}  \equiv A^0 \equiv \sqrt{\left|\frac{8}{k c^2 \alpha}\right|} \, \bar{B} - \frac{2 b}{c} \, \hat{\zeta}^0_\ell \,  B^\ell, \quad\quad A^1 \equiv \frac{\hat{\zeta}^1_\ell \, B^\ell}{\sqrt{c}}, \quad\quad A^2 \equiv \frac{\hat{\zeta}^2_\ell \, B^\ell}{\sqrt{c}}, \quad\quad \Omega  \equiv \hat{\zeta}^0_\ell \, B^\ell .
\ee

Notice that by picking these vectors to give generators in different conjugacy classes of $SL(2)$ we can change the signature in the $A^1, A^2$ space. This is a simple way of considering the extension of the $SO(2)$ symmetry in Warped Geometry to $SO(1,1)$ and even a degenerate case where one of the vectors ends up being light-like as we will see below. Let's see how this works.

Notice that the natural metric on $SL(2)$ induces a metric on the fields $A^0$, $A^1$ and $A^2$  since $\bar{B}$ has no non trivial commutator with the $SL(2)$ generators as there is no mixed term in the action (\ref{csa}). This metric is given by

\be
M_{I J} = \zeta^\ell_I \, g_{\ell k} \, \zeta^k_J
\ee

\noindent where $g_{00}=\frac{1}{2}$, $g_{+-}=g_{-+}=-1$ and the other components vanish.

It is then possible to define an $SL(2)\times U(1)$ invariant quadratic form:
\be
ds^2 = A^I M_{I J} A^J
\ee

Let us now show that this reproduces the metric of Warped $AdS_3$ space-times of all kinds and warpings, following the classification of  \cite{Anninos:2008fx}.  We have basically 3 different options. We can pick $\zeta_0$ to parameterize an elliptic (space-like) , parabolic (light-like) or hyperbolic (time-like) generator of $SL(2)$. Let's work out each case in detail. We pick the following notation for our vectors: $\zeta=( 0, +, -)$.

\begin{itemize}
\item Time-like: $\zeta_0= (0, 1 , 1)$. A convenient basis for the other generators is $\zeta_1= (0,1,-1)$, $\zeta_2=(1,0,0)$. This corresponds to the same conjugacy class as the original identification (\ref{ident}). We want to construct vacuum solutions, so following \cite{Detournay:2012pc} we set $T= -\frac{c}{24} = -\frac{k}{4}$ and $P=0$ in (\ref{bfields}). We also introduce the coordinates $r=\rho+\log{2}$ and $y=-x$. Then:
\be
ds^2 = \frac{\ell^2}{\nu^2 + 3} \left( dr^2 +  \cosh^2 r dy^2 -\frac{4 \nu^2}{\nu^2 +3} \left(dt +\sinh{r} \, dy \right)^2  \right)
\ee
\noindent where we have written
\be\label{wadscs}
\frac{1}{2 c} = \frac{\ell^2}{\nu^2 +3}, \quad\quad b^2 = \frac{ \nu^2}{2 \ell^2}, \quad\quad \left| \alpha \right| =\frac{8}{k b^2} .
\ee
This is the exact form of the time-like Warped $AdS_3$ space-times as seen in \cite{Anninos:2008fx}. Notice $c$ and $b$ give the $AdS$ radius and the warping parameter. The value of $\alpha$ can always be changed by a coordinate redefinition of $t$. All these variables are not physical in our case and determine just a particular way of choosing geometric vielbeins from the Chern-Simons variables.  
\item Space-like: $\zeta_0= (0, 1 , -1)$. A convenient basis for the other generators is $\zeta_1= (0,1,1)$, $\zeta_2=(1,0,0)$. Here, as we change the signature we need to revert the sign of the currents, so we choose $T=\frac{c}{24}= \frac{k}{4}$ and $P=0$. Using the same coordinates as above we obtain
\be
ds^2 = \frac{\ell^2}{\nu^2 + 3} \left( dr^2 - \cosh^2 r dy^2 +\frac{4 \nu^2}{\nu^2 +3} \left(dt +\sinh{r} \, dy \right)^2  \right) .
\ee
This is space-like Warped $AdS_3$ as displayed in \cite{Anninos:2008fx}. 
\item Light-like: $\zeta_0=(0,0,1)$, $\zeta_1=(0,1,1)$, $\zeta_2=(1,0,0)$. In this limit we take a Poincare patch which forces in the vacuum $T=P=0$. The metric becomes in suitable coordinates:
\be
ds^2  =  \ell^2 \left( d\rho^2 + e^{2\rho} dx^2 + e^{\rho} dx dt\right) .
\ee
This is null Warped $AdS_3$ \cite{Anninos:2008fx}. Notice that the above result requires $b = \frac{\sqrt{c}}{2}$. According to (\ref{wadscs}) this implies $\nu^2=1$ in this case. This agrees with the results in \cite{Anninos:2008fx}.

\end{itemize} 

We have thus obtained all Warped $AdS_3$ solutions within one single Chern-Simons bulk theory. The distinction between them amounts to different identifications of vielbein in only one theory. It should be pointed out that while the definition of a boundary for the space-times above is not completely clear as different components of these metrics scale with different powers, the Chern-Simons description already classifies fields according to their weight. No such ambiguity arises, thus, in the Lower Spin Gravity formulation.

\section{ Conclusions}\label{conclu}

In order to understand the basic principles behind holography, it is crucial to be able to extend the dictionary to non-$AdS$ space-times. Furthermore, as it has become clear through the study of higher-spin theories, it is sometimes useful to generalize the concept of geometry in order to write bulk theories that can describe the physics of theories with exotic symmetry groups. In parallel to these developments in holography, it is always important to understand and extend the space of QFTs that can be solved exactly. They gives us a benchmark on what to expect from strongly coupled field theories and they represent a lamppost to develop and check holographic dualities.

In the present work we have presented results that sit at the intersection of these roads. First, we have presented a physical definition of a WCFT by highlighting the important role of the boost symmetry $t \rightarrow t+ v x$ . It turns out that this singles out WCFTs among scaling theories in much the same ways as Lorentz symmetry singles out CFTs. With this new insight we developed the necessary geometric setting to couple WCFTs to background fields. This geometry is not Riemannian. It carries the boost symmetry action in its tangent space instead of the Lorentz algebra. Once the dust settles, in generic dimension $d$, this geometry can be described by a diffeomorphic covariant language where some components of the torsion vanish (as in Riemannian geometry) but also some components of the curvature vanish (as in Weitzenb\"{o}ck geometry). We derived these conditions by demanding that the geometry supports a scaling structure. Physically, this means that there exists preferred axes that can be assigned scaling weights if one so chooses. The role of this geometric construction is analogous to the existence of a complex structure in Riemannian geometry. This allows for the coupling of a WCFT to the geometry.

These developments allowed us to build a fully covariant formalism for WCFTs in curved spaces. With it we can study the symmetries of the problem and the associated conserved currents by using background field methods. These methods are powerful as they imply certain invariances of the partition functions of these theories under Warped Weyl scalings (\ref{warpedweyl}). WCFTs do not couple to a metric naturally, so their symmetries are better represented by these new transformations acting on background fields $A_\mu$ and $\bar{A}_\mu$ in $d=2$.

An important upshot of having a fully covariant formalism is that we could write explicitly two examples of (free) WCFTs. This is important, as we now expect to be able to extend this family in an analogous way to what is done in  the existing literature on CFTs. This could be interesting as some specific examples could prove useful both in condensed matter physics applications, where it seems natural to give up Lorentz symmetry, and from a purely formal point of view. For example, it is expected that there exist Warped Minimal Models that can be written explicitly. One reason to suspect this is the case is that there is a strong connection between WCFTs and chiral CFTs. It seems that if the CFT possess a target space dimension where physics is manifestly local, WCFT makes this manifest by putting target space symmetries together with base space symmetries. As such, WCFTs treat democratically spin 1 and spin 2 currents from the point of view of the chiral CFT.

This last point is of importance. Encouraged by the success in unifying the discussion of currents of different spins one could conjecture the existence of a similar formalism that could be developed in the study of $W_N$ CFTs. One of the main difficulties in understanding deformations of $W_N$ models is that higher spin currents are irrelevant under the usual renormalisation group classification. Deformations of this sort have been recently considered in \cite{deBoer:2014fra}.  This classification, however, singles out the scaling generator in the $W_N$ algebra and breaks democracy in a manifest way. If one had a more covariant formalism where conformal and higher spin symmetries were discussed within the same framework it might be possible to keep these deformations under control.

In section \ref{holsec} we discussed a general holographic construction of bulk duals. This construction is minimal in the sense that it does not include any bulk fields that are not required by symmetry. This is certainly not the case of the Topologically Massive Gravity and massive vector model constructions. This minimal bulk theory can be written as a $SL(2) \times U(1)$ Chern-Simons theory. We call this theory Lower Spin Gravity, in analogy to the Chern-Simons formulation of Higher Spin Gravity. In order to arrive at this construction it was critical to have a precise understanding of the boundary warped geometry. We also discussed how to obtain the infinite dimensional symmetry algebra by adapting the usual discussions for $AdS_3$ in the $SL(2)$ Chern-Simons description. It is important to stress that the $U(1)$ plays a geometric role, exactly as $SL(2)$ does. This introduces non standard boundary terms in defining the Chern-Simons theory. Lastly we were able to connect this formalism with the usual Warped $AdS_3$ space-times discussed in the literature \cite{Anninos:2008fx,Anninos:2008qb}.

As a future direction, we mention that one could use Lower Spin Gravity to better understand holographic renormalization in this context. This theory separates clearly background fields with different scaling weight as opposed to the usual Warped $AdS_3$ descriptions. This simplifies greatly the definition of a boundary and could help understand the necessary counter-terms. Furthermore, Lower Spin Gravity is a much simpler and better behaved theory than Topologically Massive Gravity.

Another possible direction consists in using the formalism developed here coupled to what is known in generic Chern-Simons theories \cite{Ammon:2013hba,deBoer:2013vca} to compute entanglement entropy both in a QFT and a holographic setup. Notice that the knowledge of how partition functions transform under changes of the background sources makes this problem very similar to existing CFT calculations. From the point of view of the bulk, the understanding of concepts analogous to geodesics \cite{Anninos:2013nja} and minimal surfaces in Warped Geometry allows for the generalization of standard holography techniques. This is work in progress \cite{inprog}.

\section*{Acknowledgements}
We are delighted to acknowledge enlightening discussions with H. Afshar, J. de Boer, A. Castro, N. Iqbal and especially J. Jottar. BR acknowledges support from the Swiss National Science Foundation through the fellowship PBBEP2 144805.

\pagebreak



\end{document}